\documentclass[twocolumn,superscriptaddress,preprintnumbers,amsmath,amssymb,prd]{revtex4}

\usepackage{graphicx}
\usepackage{dcolumn}
\usepackage{bm}
\usepackage{indentfirst} 
\usepackage{color}
\usepackage{CJK}
\usepackage{booktabs}
\usepackage{float}
\usepackage[colorlinks,linkcolor=blue,urlcolor=blue,citecolor=blue]{hyperref}

\usepackage[colorlinks,linkcolor=blue,urlcolor=blue,citecolor=blue]{hyperref}

\usepackage[colorlinks,linkcolor=blue,urlcolor=blue,citecolor=blue]{hyperref}

\usepackage{tikz,xcolor,hyperref}

\definecolor{lime}{HTML}{A6CE39}
\DeclareRobustCommand{\orcidicon}{
	\begin{tikzpicture}
	\draw[lime, fill=lime] (0,0) 
	circle [radius=0.16] 
	node[white] {{\fontfamily{qag}\selectfont \tiny ID}};
	\draw[white, fill=white] (-0.0625,0.095) 
	circle [radius=0.007];
	\end{tikzpicture}
	\hspace{-2mm}
}
\foreach \x in {A, ..., Z}{%
	\expandafter\xdef\csname orcid\x\endcsname{\noexpand\href{https://orcid.org/\csname orcidauthor\x\endcsname}{\noexpand\orcidicon}}
}
\foreach \x in {A, ..., Z}{%
	\expandafter\xdef\csname orcid\x\endcsname{\noexpand\href{https://orcid.org/\csname orcidauthor\x\endcsname}{\noexpand\orcidicon}}
}

\begin{document}
\begin{CJK*}{UTF8}{gbsn}

\title{A Novel Framework for Characterizing Spacetime Microstructure with Scaling}

\author{Weihu Ma(马维虎)\orcidA{}}
    \email[Correspondence email address: ]{maweihu@fudan.edu.cn}
\author{Yu-Gang Ma(马余刚)\orcidB{}}
    \email[Correspondence email address: ]{mayugang@fudan.edu.cn}
    \affiliation{Key Laboratory of Nuclear Physics and Ion-beam Application (MOE), Institute of Modern Physics, Fudan University, Shanghai 200433, China}
    \affiliation{Shanghai Research Center for Theoretical Nuclear Physics, NSFC and Fudan University, Shanghai 200438, China}

\date{\today} 

\begin{abstract}
The study of physics at the Planck scale has garnered significant attention due to its implications for understanding the fundamental nature of the universe. At the Planck scale, quantum fluctuations challenge the classical notion of spacetime as a smooth continuum, revealing a complex microstructure that defies traditional models. This study introduces a novel scaling-based framework to investigate the properties of spacetime microstructures. By deriving a scaling-characterized metric tensor and reformulating fundamental equations—including the geodesic, Einstein field, Klein-Gordon, and Dirac equations—into scaling forms, the research reveals new properties of local spacetime dynamics. Remarkably, the golden ratio emerges naturally in linear scale measurements, offering a potential explanation for the role of the Planck length in resolving ultraviolet (UV) divergence. Furthermore, the study demonstrates how scale invariance in spacetime can restore classical geometric stability through the renormalization group equations. These findings significantly revise classical geometric intuitions, providing a fresh lens for understanding quantum fluctuations and offering promising insights for advancing quantum gravity theories.
\end{abstract}


\maketitle

\section{Introduction}
The Planck length is particularly significant, as it is the scale at which quantum gravitational effects are expected to become apparent, thus interactions require a working theory of quantum gravity to be studied. It is also a natural estimate for the fundamental string length scale and the characteristic size of compact extra dimensions \cite{bibitem1}. At the Planck scale, current models fail to describe the universe effectively, and a scientific model to explain the behavior of the physical universe has yet to be established. Our current understanding suggests that the universe began $\sim 10^{-43}$ seconds after the Big Bang. At this scale, the uncertainty principle reveals the quantum fluctuations of spacetime, which are integral to elucidating the original structure of the universe, the center of a black hole, and the structure of nature at a very short scale, in a framework of quantum gravity \cite{bibitem2}. 

Many studies seek to address these challenges in quantum gravity, including two of the most potential of them, i.e., Loop Quantum Gravity (LQG) and String Theory. LQG is an approach to quantum gravity that quantizes spacetime itself by describing it in terms of discrete loops or spin networks \cite{bibitem2, Gambini, Ashtekar, Thiemann, Pullin}. It provides a non-perturbative and background-independent framework, where space is quantized, and the geometry emerges from the quantum states of the gravitational field. LQG introduces a new set of variables and applies loop quantization techniques, leading to finite areas and volumes, and avoiding singularities such as those in black holes  \cite{Gambini2}. String Theory posits that the fundamental constituents of the universe are not point particles, but rather one-dimensional “strings” that vibrate at different frequencies to produce different particles. It provides a unified framework for all fundamental forces, including gravity, by incorporating the concept of extra dimensions and supersymmetry \cite{Zwiebach, bibitem1, LeBihan}. String Theory has successfully incorporated gravity into quantum mechanics, offering a potential solution to the problem of quantum gravity and explaining the existence of different particle types through string vibrations \cite{Vaid}. Recent research in quantum gravity reveals various new approaches and theories \cite{Klauder,Oriti,Berenstein,Pullin,Shojai,Carlip,Harlow,Pawlowski,Braunstein,Giesel,King,Bajardi,Chakraborty,Saueressig,Bianchi}. Although these studies have greatly improved our understanding of physics at the Planck scale, they are obviously not exhaustive.

Due to its incompleteness, quantum gravity imposes constraints on our understanding of spacetime at the smallest scales. This challenge has piqued the interest of many researchers who are drawn to the field in pursuit of unraveling its mysteries. At this scale, spacetime is hypothesized to deviate from the classical notion of a smooth continuum, potentially exhibiting a foam-like, fluctuating structure \cite{bibitem3,bibitem4}. This concept is rooted in the idea that the Heisenberg uncertainty principle could give rise to microscopic irregularities, as postulated by Wheeler. Additionally, Tryon's hypothesis suggests that our universe may have originated from a quantum vacuum fluctuation, similar to the Big Bang scenario \cite{bibitem5}. Quantum fluctuations are not just theoretical curiosities but integral to the fabric of the universe's microstructure. The significance of these fluctuations is underscored by ongoing research that highlights the necessity for a deeper exploration into the foundational characteristics of spacetime at the Planck scale. This quest for understanding is driven by the goal of reconciling the principles of quantum mechanics with the geometry of spacetime, with the ultimate goal of revealing the true nature of reality at its most fundamental level.

The Planck scale represents a frontier in theoretical physics, where quantum fluctuations dominate, and classical descriptions of spacetime become inadequate. The present work aims to explore the microstructure of spacetime at the Planck scale through micro-measurements of spacetime characterized by scaling functions. By examining the scaling behavior of micro-geometry, the research aims to provide a new perspective for exploring the nature of spacetime microstructures and offer valuable insights into fundamental spacetime properties. We develop a novel scaling-characterized metric tensor derived from the Lorentz scalar line element, enabling the analysis of spacetime microstructure. The differential operators are transformed into scaling form, providing a mathematical framework for examining fluctuation properties. The content is structured as follows:
II. Micro-measuring principle and Scaling Characterization;
III. Scaling Characterization in Metric;
IV. The Lorentz Factor at the micro-scale;
V. Micro Measurement in Spacetime with Schwarzschild Metric;
VI. Scaling Geodesic Equations and Einstein Field Equations;
VII. The Golden Ratio at the micro-scale;
VIII. Scaling Klein-Gordon Equation and Dirac Equation;
IX. Summary.

\section{Micro-measuring Principle and Scaling Characterization}

Heisenberg's principle suggests that spacetime exhibits quantum fluctuations on a microscopic scale. These fluctuations result in uncertainties in the positions of points within spacetime coordinates and cause variations in microlengths. At the Planck scale, traditional concepts of spacetime geometry become inadequate due to quantum fluctuations. These fluctuations imply that spacetime is not a smooth manifold but has a more intricate structure.

We model the spacetime fluctuations using the differential \( dx^{\alpha} \), representing spacetime microelements as microlength measurements relative to a reference length \( dx^{\alpha}_{0} \). Here, $\alpha$ (equal to 0, 1, 2, 3, ...) denotes the coordinate index in orthonormal frames. The microelements represent small segments of spacetime, making it a natural choice for modeling fluctuations at quantum scales, where spacetime might exhibit complex, non-continuous behavior. In quantum mechanics, uncertainty and fluctuations are inherent due to the Heisenberg uncertainty principle. Modeling spacetime at quantum scales involves accounting for these uncertainties, which microelements naturally accommodate by allowing for the representation of small changes and variations. Defining spacetime quantum fluctuations using differentials \( dx^{\alpha} \) is a rational choice that fits well with current theoretical frameworks in physics. It allows for a coherent description of spacetime at very small scales, integrating principles of general relativity and quantum mechanics. This definition provides a useful starting point for exploring the quantum nature of spacetime.

The measurements of \( dx^{\alpha} \) relative to \( dx^{\alpha}_{0} \) is introduced by arbitrary scaling functions:
\begin{equation}
r_{l^{\alpha}} = \frac{dx^{\alpha}}{dx^{\alpha}_{0}} = L^{\alpha}(X^{\alpha}(\tau)).
\end{equation}
These scaling functions $L^{\alpha}(X^{\alpha}(\tau))$ suggest a way to quantify these quantum fluctuations, denoted by spacetime scales \(X^{\alpha}\), and quantify how the microlength measurements deviate from their reference lengths due to quantum fluctuations. \( dx^{\alpha}_{0} \) are micro reference lengths, serving as the baseline measurements in the absence of fluctuations. It establishes a well-defined reference scale that allows us to measure the deviations. To measure the micro proper length, scaling function $\tau$(X(s)) is introduced and defined as
\begin{equation}
\begin{aligned}
r_{l}=\frac{d\lambda}{d\lambda_{0}}=\tau(X(s))
\end{aligned}
\end{equation}
in a similar way, where $d\lambda$ is the proper length microelement, $d\lambda_{0}$ is the reference length, and $s$ is defined as the proper time scale.

Consequently, the measurements of $dx^{\alpha}$ relative to proper length $d\lambda$ is given by
\begin{equation}
\begin{aligned}
\frac{dx^{\alpha}}{d\lambda}=\frac{L^{\alpha}(X^{\alpha}(\tau))}{\tau(X(s))} \cdot \frac{dx^{\alpha}_{0}}{d\lambda_{0}}=\frac{L^{\alpha}(X^{\alpha}(\tau))}{\tau(X(s))},
\end{aligned}
\end{equation}
which involves choosing consistent reference lengths for $dx^{\alpha}_{0}$ = $d\lambda_{0}$. The Planck length is considered the optimal reference length as it represents the smallest measurable length scale in quantum gravity, making it a natural choice for investigating spacetime at this scale.

With these definitions, there are three types of spacetime micro measurements:

(a), When $L^{\alpha}=constant$ ($\tau=constant$), indicating that micro length measurements are conducted in a static mode (trivial measurement) without fluctuations, independent of $X^\alpha$($X$), indicating a perfectly stable spacetime region. When measurements are performed exclusively along a single dimension, the non-fluctuating mode can be represented as $L^{\alpha}=1$ as the $constant$ can be incorporated into the reference length through appropriate scaling. 

(b), When $L^{\alpha}=K^{\alpha}_{1}X^{\alpha}+K^{\alpha}_{2}$ ($\tau=K_{1}X+K_{2}$), micro length measurements are generally performed in a linear fluctuating mode, where $K^{\alpha}(K)$ are coefficients. This suggests a linear relation between the microlength and their fluctuation. When measurements are performed exclusively along a single dimension, the linear fluctuating mode can be represented as $L^{\alpha}=X^{\alpha}$ as the coefficients can be incorporated into the reference length through translation and scaling.

(c), When $L^{\alpha}=L^{\alpha}(X^{\alpha})$ ($\tau=\tau(X)$) are nonlinear functions, micro length measurements occur in a nonlinear fluctuating mode.

Micro measurements might involve complex scaling functions, or even discrete rather than continuous measurements, leading to distinct spacetime structures. The exploration of these forms is crucial for a more comprehensive view of spacetime microstructure.

In this framework, spacetime fluctuation measurement can be described by the scaling function with non-trivial measurement. Additionally, the fluctuation induced by $X^\alpha$ signifies micro structures along the $\alpha$-axis. If the given $L^{\alpha}$ is the same for all axes, it indicates an isotropic structure. 

Define $d\bar{x}^{\alpha}=\bar{L}^{\alpha}dx^{\alpha}_{0}=L^{\alpha}(\zeta^{\alpha}X^{\alpha})dx^{\alpha}_{0}$ with $\zeta^{\alpha}(\tau)$ being the $\alpha-axis$ rescaling factor of scale transformation. And also define $d\bar{\lambda}=\bar{\tau}d\lambda_{0}=\tau(\chi X)d\lambda_{0}$ with $\chi(s)$ being the rescaling scale transformation on $X(s)$. Then we can get
\begin{equation}
\begin{aligned}
\frac{d^{2}x^{\alpha}}{d\lambda^{2}}&=\frac{d}{d\lambda}(\frac{dx^{\alpha}}{d\lambda})=\frac{1}{d\lambda}(\frac{d\bar{x}^{\alpha}}{d\lambda}-\frac{dx^{\alpha}}{d\lambda})\\
&=\frac{\bar{L}^{\alpha}-L^{\alpha}}{\tau^{2}}\frac{1}{d\lambda_{0}},
\end{aligned}
\end{equation}
and the change of $r_{l^{\alpha}}$ relative to $r_{l}$ is
\begin{equation}
\begin{aligned}
\frac{dr_{l^{\alpha}}}{dr_{l}}&=\frac{d\bar{x}^{\alpha}-dx^{\alpha}}{d\bar{\lambda}-d\lambda}=\frac{\bar{L}^{\alpha}-L^{\alpha}}{\bar{\tau}-\tau}.
\end{aligned}
\end{equation}

From $d(dx^{\alpha})=d\bar{x}^{\alpha}-dx^{\alpha}=(\frac{d\bar{x}^{\alpha}}{dx^{\alpha}}-1)dx^{\alpha}=(\frac{\bar{L}^{\alpha}}{L^{\alpha}}-1)dx^{\alpha}$ and $d(dx^{\alpha})=\frac{dL^{\alpha}}{dX^{\alpha}}dX^{\alpha}dx_{0}^{\alpha}$, we get 
\begin{equation}
\begin{aligned}
&\frac{dX^{\alpha}}{dx^{\alpha}}=\hat{a}_{\alpha}\frac{1}{dx_{0}^{\alpha}}
\end{aligned}
\end{equation}
with 
\begin{equation}
\begin{aligned}
&\hat{a}_{\alpha}=\frac{\bar{L}^{\alpha}/L^{\alpha}-1}{dL^{\alpha}/dX^{\alpha}},
\end{aligned}
\end{equation}
thus we have the first and second-order partial differential operators transformed in scaling form by
\begin{equation}
\begin{aligned}
&\frac{\partial}{\partial x^{\nu}}=\frac{dX^{\nu}}{dx^{\nu}}\frac{\partial }{\partial X^{\nu}}=\frac{1}{dx_{0}^{\nu}}\hat{a}_{\nu}\frac{\partial}{\partial X^{\nu}},
\end{aligned}
\end{equation}
\begin{equation}
\begin{aligned}
\frac{\partial^{2} }{\partial x^{\mu}\partial x^{\nu}}&=\frac{dX^{\mu}}{dx^{\mu}}\frac{\partial}{\partial X^{\mu}}(\frac{\partial}{\partial x^{\nu}})=\frac{1}{dx_{0}^{\mu}}\frac{1}{dx_{0}^{\nu}}\\
&\cdot(\hat{a}_{\mu}\hat{a}_{\nu}\frac{\partial^{2}}{\partial X^{\mu}\partial X^{\nu}}-\hat{a}_{\mu}\hat{b}_{\nu}\frac{\partial X^{\nu}}{\partial X^{\mu}}\frac{\partial}{\partial X^{\nu}}),
\end{aligned}
\end{equation}
where
\begin{equation}
\begin{aligned}
&\hat{b}_{\nu}=-\frac{d\hat{a}_{\nu} }{dX^{\nu}}.
\end{aligned}
\end{equation}
Eq. (8) and (9) connect with the differential operators to scaling functions, which are introduced to describe how the fluctuation of spacetime microelement can be adjusted according to a scaling factor $\zeta^{\nu}$. These equations introduce the factors $\hat{a}_{\nu}$ and $\hat{b}_{\nu}$ that modify the behavior of differential operators to account for changes in scale. 

Similarly, introduce a scale transformation factor $\chi$ acting on $X(s)$. And from $d(d\lambda)=d\bar{\lambda}-d\lambda=(\frac{\bar{\tau}}{\tau}-1)d\lambda$ and $d(d\lambda)=\frac{d\tau}{dX}dXd\lambda_{0}$, we can get $\frac{dX}{d\lambda}=\hat{a}\frac{1}{d\lambda_{0}}$. Thus, we get the first and second-order differential operators transformed in scaling form by
\begin{equation}
\begin{aligned}
&\frac{d}{d\lambda}=\frac{dX}{d\lambda}\frac{d}{dX}=\frac{1}{d\lambda_{0}}\hat{a}\frac{d}{dX},
\end{aligned}
\end{equation}
\begin{equation}
\begin{aligned}
&\frac{d^{2}}{d\lambda^{2}}=\frac{dX}{d\lambda}\frac{d}{dX}(\frac{d}{d\lambda})=\frac{1}{[d\lambda_{0}]^{2}}(\hat{a}^{2}\frac{d^{2}}{dX^{2}}-\hat{a}\hat{b}\frac{d}{dX}),
\end{aligned}
\end{equation}
where $\hat{a}=\frac{\bar{\tau}/\tau-1}{d\tau/dX}$ and $\hat{b}=-\frac{d\hat{a}}{dX}$.

Introducing the scale transformation $\bar{X}^{\alpha}=\zeta^{\alpha}X^{\alpha}$ with rescaling factor $\zeta^{\alpha}=\zeta^{\alpha}(X^{\alpha})$, we obtain $\frac{d\bar{X}^{\alpha}}{dX^{\alpha}}=\zeta^{\alpha}+X^{\alpha}\frac{d\zeta^{\alpha}}{dX^{\alpha}}$ and $\frac{dX^{\alpha}}{d\bar{X}^{\alpha}}=\frac{1}{\zeta^{\alpha}}$. Subsequently, this leads to the derivation of
\begin{equation}
\begin{aligned}
&\frac{\partial}{\partial \bar{X}^{\alpha}}=\frac{dX^{\alpha}}{d\bar{X}^{\alpha}}\frac{\partial}{\partial X^{\alpha}}=\frac{1}{\zeta^{\alpha}}\frac{\partial}{\partial X^{\alpha}};
\end{aligned}
\end{equation}
\begin{equation}
\begin{aligned}
&\frac{\partial^{2}}{\partial \bar{X}^{\beta}\partial \bar{X}^{\alpha}}=\frac{dX^{\beta}}{d\bar{X}^{\beta}}\frac{\partial}{\partial X^{\beta}}(\frac{\partial}{\partial \bar{X}^{\alpha}})\\
&=\frac{1}{\zeta^{\beta}}\frac{1}{\zeta^{\alpha}}\frac{\partial^{2}}{\partial X^{\beta}\partial X^{\alpha}}-\{\frac{1}{\zeta^{\beta}[\zeta^{\alpha}]^{2}}\frac{d\zeta^{\alpha}}{dX^{\alpha}}\}\frac{\partial X^{\alpha}}{\partial X^{\beta}}\frac{\partial}{\partial X^{\alpha}}.
\end{aligned}
\end{equation}
Eq. (13) and (14) use the idea of scale transformations by focusing on partial derivatives of scaled coordinates. These equations describe how spacetime coordinates are rescaled using the factor $\zeta^{\alpha}$. This rescaling is crucial for analyzing how measurements adapt in transitioning between a linear and a nonlinear scaling regime when considering together Eq. (8) and (9), which will be further explained in the following sections VI-VIII.

Similarly, the transformation can be introduced for $\bar{X}=\chi X$ with rescaling factor $\chi=\chi(X)$. Thus $\frac{d\bar{X}}{dX}=\chi+X\frac{d\chi}{dX}$ and $\frac{dX}{d\bar{X}}=\frac{1}{\chi}$. Subsequently, we obtain
\begin{equation}
\begin{aligned}
&\frac{d}{d\bar{X}}=\frac{dX}{d\bar{X}}\frac{d}{dX}=\frac{1}{\chi}\frac{d}{dX};
\end{aligned}
\end{equation}
\begin{equation}
\begin{aligned}
&\frac{d^{2}}{d\bar{X}^{2}}=\frac{dX}{d\bar{X}}\frac{d}{dX}(\frac{d}{d\bar{X}})=\frac{1}{[\chi]^{2}}\frac{d^{2}}{dX^{2}}-\frac{1}{[\chi]^{3}}\frac{d\chi}{dX}\frac{d}{dX}.
\end{aligned}
\end{equation}

The introduction of these transformations establishes a mathematical framework for investigating the nature of spacetime at the micro-scale, which is further explored in the subsequent sections.

Micro-measuring principle at micro-scale: the distance between two infinitesimally close points in spacetime is denoted by $dx$, and its distance measurement requires definition through a comparison with a reference length. If $dx$ remains unchanging, the measured value only relies on the chosen reference length. We can establish a consistent spacetime measurement and facilitate comparisons of their respective lengths by selecting an appropriate fixed-length reference for measuring the distance. In cases where the micro distance $dx$ in spacetime undergoes fluctuations for some reason, the measurement of $dx$, relative to the reference length, is introduced as the scaling function. In the framework of quantum gravity theory, the reason for the fluctuations of spacetime microelements is considered to come from the quantum effects of spacetime at the microscopic scale, which can be inferred through the Heisenberg uncertainty principle. The measurement of microelements in quantum gravitational spacetime can be defined as fluctuating, including linear and nonlinear fluctuations. The spacetime microelement, defined by scaling functions, allows for the exploration of spacetime microstructure. This conceptualization enables the development of new mathematical tools and theories to investigate the nature of spacetime at the micro-scale.

\section{Scaling Characterization in Metric}
The Lorentz scalar line element, also known as the spacetime invariant interval, is a fundamental concept in the theory of relativity. It measures the infinitesimal distance between two events in spacetime. It is significant due to its invariance under Lorentz transformations, meaning it remains constant for all observers, regardless of their relative motion. This invariance is a cornerstone of Einstein's theory of relativity, ensuring that the laws of physics are the same in all inertial frames of reference.

The Lorentz scalar line element is crucial for maintaining the consistency of physical laws across all scales. It provides a foundational framework for understanding both macroscopic and microscopic phenomena, preserving spacetime's geometric and causal properties across different observational perspectives. Kothawala \cite{Kothawala} introduces the concept of a "zero-point" length, suggesting that even at the smallest scales, the Lorentz scalar line element plays a significant role in defining the geometric properties of spacetime. As described by Volovik \cite{Volovik}, it is linked to fundamental constants like the Planck length, acting as a natural "zero-point" length. This connection is vital for exploring theories that aim to unify general relativity and quantum mechanics at the Planck scale. The Lorentz scalar line element helps characterize quantum fluctuations by providing a consistent metric that remains invariant under transformations, aiding in the interpretation of quantum behaviors.

The most general form of the Lorentz scalar line element is given by
\begin{equation}
\begin{aligned}
d\lambda^{2}&=g_{\alpha\beta}dx^{\alpha}dx^{\beta},
\end{aligned}
\end{equation}
where $g_{\alpha\beta}$ is the metric tensor, and the Einstein notation is used. The potential fluctuating nature of spacetime at small scales can be studied using the invariant interval. Utilizing the scaling representation defined in previous equations, we derive an equation for the metric tensor related to micro-measurements of spacetime fluctuations
\begin{equation}
\begin{aligned}
\tau^{2}(s)d\lambda^{2}_{0}&=g_{\alpha\beta}L^{\alpha}(\tau)L^{\beta}(\tau)dx^{\alpha}_{0}dx^{\beta}_{0}
\end{aligned}
\end{equation}
with Einstein notation used. When $\tau=\tau_{C}$, $L^{\alpha}=L^{\alpha}_{C}$, and $L^{\beta}=L^{\beta}_{C}$ are $Constants$ respectively, we define a static measurement as
\begin{equation}
\begin{aligned}
\tau_{C}^{2}&=g_{\alpha\beta}L^{\alpha}_{C}L^{\beta}_{C},
\end{aligned}
\end{equation}
in a consistent choice of $d\lambda_{0}=dx^{\alpha}_{0}=dx^{\beta}_{0}$. This indeed corresponds to the definition of the classical Lorentz scalar line element. Consequently, for a non-static measurement, we have
\begin{equation}
\begin{aligned}
\tau^{2}(X(s))&=g_{\alpha\beta}L^{\alpha}(X^{\alpha}(\tau))L^{\beta}(X^{\beta}(\tau)),
\end{aligned}
\end{equation}
with the consistent choice of $d\lambda_{0}=dx^{\alpha}_{0}=dx^{\beta}_{0}$.

Define the micro measurement of proper time to be $r_{t}=\frac{dt}{dt_{0}}=r_{t}(s)$ with reference time length $dt_{0}$. Here $s$ is the defined proper time scale. Introduce light speed measurement of $\frac{d\lambda}{dt}=r_{c}\frac{d\lambda_{0}}{dt_{0}}=r_{c}(\nu)\frac{d\lambda_{0}}{dt_{0}}$ from scale transformation of $\tau=r_{l}=r_{c}\cdot r_{t}$. $\nu=\nu(s)$ is defined as speed scale and a function of $s$. Thus we can get
\begin{equation}
\begin{aligned}
r_{c}^{2}&=g_{\alpha\beta}\frac{L^{\alpha}L^{\beta}}{r_{t}^{2}}
\end{aligned}
\end{equation}
with Einstein notation.

Define proper speed measurements from $r_{v^{\alpha}}=\frac{r_{l^{\alpha}}}{r_{t}}=\frac{L^{\alpha}}{r_{t}}=U^{\alpha}(V^{\alpha}(s))$.  
For space components, $r_{v^{i}}=U^{i}(V^{i}(s))=\frac{r_{l^{i}}}{r_{t}}=\frac{r_{l^{i}}}{r_{c}r_{T}}\frac{r_{c}r_{T}}{r_{t}}=\frac{v^{i}}{r_{c}}\frac{r_{c}r_{T}}{r_{t}}=\frac{v^{i}}{r_{c}}\frac{r_{l^{0}}}{r_{t}}=\frac{v^{i}}{r_{c}}r_{v^{0}}$, where $r_{c}r_{T}=r_{l^{0}}=r_{v^{0}}r_{t}$ and then the coordinate speed measurement $v^{i}=\frac{r_{l^{i}}}{r_{T}}$ with $r_{T}$ being the measurement of the coordinate time micro interval. Using Eq. (21), Thus we have
\begin{equation}
\begin{aligned}
r_{c}^{2}&=r_{c}^{2}(\nu(s))=g_{\alpha\beta}U^{\alpha}(V^{\alpha}(s))U^{\beta}(V^{\beta}(s))
\end{aligned}
\end{equation}
with Einstein notation.

Performing differentiation on both sides of Eq. (20) with respect to $s$, we aim to find how the metric tensor and scaling components evolve as time scale $s$ changes. Thus, we obtain
\begin{equation}
\begin{aligned}
0&=\frac{d\tau}{ds}(\frac{dg_{\alpha\beta}/d\tau}{g_{\alpha\beta}}-\frac{2}{\tau}+\frac{dL^{\alpha}/d\tau}{L^{\alpha}}+\frac{dL^{\beta}/d\tau}{L^{\beta}})g_{\alpha\beta}L^{\alpha}L^{\beta}
\end{aligned}
\end{equation}
with Einstein notation. This equation governs how the metric tensor evolves under scaling transformations, showing the intricate dependencies between derivatives of scale variables and the metric. The terms represent a balance between metric changes and scaling influences, indicating that the metric must adjust to maintain consistency in spacetime geometry under scaling. 

Under conditions of $L^{\alpha}\neq0$, $L^{\beta}\neq0$, and $d\tau/ds\neq0$, we get equations for cases of $g_{\alpha\beta}\neq0$
\begin{equation}
\begin{aligned}
\frac{dg_{\alpha\beta}/d\tau}{g_{\alpha\beta}}&=\frac{2}{\tau}-\frac{dL^{\alpha}(\tau)/d\tau}{L^{\alpha}(\tau)}-\frac{dL^{\beta}(\tau)/d\tau}{L^{\beta}(\tau)}.
\end{aligned}
\end{equation}
The conditions $L^{\alpha}\neq0$ and $L^{\beta}\neq0$ ensure that the microelements being measured are non-zero on the chosen scale. Such constraints help prevent ultraviolet (UV) divergence, ensuring the physical validity of the model \cite{bibitem1, bibitem2}. $d\tau/ds \neq 0$ ensures that $\tau(s)$ varies with $s$ in a non-trivial way. This condition ensures that $\tau(s)$ is not a constant function but varies with the parameter $s$. Physically, this means that the system is evolving or changing over time or space, representing a fluctuating mode in space or time scales. If $d\tau/ds = 0$, $\tau(s)$ would be constant, implying no change or evolution, which would trivialize the analysis. These conditions ensure that the metric tensor and scaling behavior described by  Eq. (24) are physically meaningful. They allow the equation to capture the dynamic, non-trivial structure of spacetime at microscopic scales, where quantum effects and scaling laws significantly influence spacetime geometry. After integral Eq. (24), we get
\begin{equation}
\begin{aligned}
g_{\alpha\beta}&= \kappa_{\alpha\beta}\frac{\tau^{2}}{L^{\alpha}L^{\beta}},
\end{aligned}
\end{equation}
where $g_{\alpha\beta}=g_{\beta\alpha}$, and $\kappa_{\alpha\beta}$ are integral constant and satisfy the relation of
\begin{equation}
\begin{aligned}
\Sigma_{\alpha,\beta}\kappa_{\alpha\beta}=1.
\end{aligned}
\end{equation}

\section{The Lorentz factor at the micro-scale}
According to the equivalence principle of general relativity \cite{bibitem8,bibitem9,bibitem10}, the influence of a gravitational field is locally indistinguishable from an inertial frame, meaning the laws of physics reduce to those of special relativity in a sufficiently small local region. This principle allows the use of a local inertial frame where gravitational effects can be approximately neglected. Consequently, in such a locally inertial frame, the metric tensor $g_{\alpha\beta}$ can be approximated by the flat spacetime metric $\eta_{\alpha\beta}$. Therefore, in a local inertial frame or even in a flat spacetime, $\kappa_{\alpha\beta}$ in a D-dimensional flat spacetime with metric $g_{\alpha\beta}=\eta_{\alpha\beta}$ can be determined. Here, $\eta_{\alpha\beta}=(\pm, \mp1, ..., \mp1)$ is the D-dimensional Minkowski spacetime metric for time-like and space-like interval \cite{bibitem11}.

In the realm of physics, special relativity \cite{bibitem12,bibitem13} provides a framework for understanding how time and space are perceived differently by observers in relative motion. The Lorentz factor is a key component of this theory, quantifying the degree of time dilation and length contraction experienced by objects moving at significant fractions of the speed of light. However, at the microscopic scale, where quantum effects play a crucial role, the influence of spacetime fluctuations on these relativistic effects becomes a fascinating subject of study \cite{Lieu1,Jacobson,Abdo,Zimdahl,Maziashvili,Lieu2}. This exploration seeks to extend the understanding of the Lorentz factor by incorporating the impact of quantum fluctuations, offering a nuanced view of how intrinsic measurements of space and time behave at the quantum level. By examining a scenario involving microscopic length measurements along a given spatial axis, we delve into the interplay between relativity and quantum mechanics, introducing scale parameters to capture the essence of intrinsic measurements in a D-dimensional space.

In special relativity, the Lorentz factor describes the relativistic effects on time and length measurements due to an object's velocity. At the micro-scale, where quantum effects become significant, we explore how spacetime fluctuations influence this factor. We consider a scenario where a microscopic length is measured at a velocity $v^{1}$ along a given spatial axis $i$-axis (i=1, 2, 3, ..., D-1, chose $i=1$). We introduce the scale parameter  $\varsigma^{i}=\sigma^{i}/\tau$ with $\sigma^{i}$ representing the micro space length measurement along the i-axis in the rest frame, i.e., representing the intrinsic measurements. Let $\tau$ denote the micro proper length measurement. Thus, the micro intrinsic measurements in a D-dimensional space can be defined as:
\begin{equation}
\begin{aligned}
&\sigma^{\alpha}=(\tau,\sigma^{i})=(\tau,\varsigma^{i}\tau).
\end{aligned}
\end{equation}
in an instantaneous rest frame. For the relativistic length contraction effect, we have:
\begin{equation}
\begin{aligned}
r_{l^{1}}&=\varsigma^{1} \tau\sqrt{1-\frac{[v^{1}]^{2}}{[v^{0}]^{2}}}
\end{aligned}
\end{equation}
with $v^{1}=\frac{r_{l^{1}}}{r_{T}}=\frac{r_{v^{1}}\cdot r_{c}}{r_{v^{0}}}=\varsigma^{1}r_{c}/\gamma_{Ltz}^{2}$ and $v^{0}=r_{c}$. For the relativistic time dilation effect, we have:
\begin{equation}
\begin{aligned}
&\frac{r_{l^{0}}}{r_{c}}=r_{t}/\sqrt{1-\frac{[v^{1}]^{2}}{[v^{0}]^{2}}}
\end{aligned}
\end{equation}
Here, the Lorentz factor $\gamma_{Ltz}=1/\sqrt{1-\frac{[v^{1}]^{2}}{[v^{0}]^{2}}}$. Thus we derive equations
\begin{equation}
\begin{cases}
\vartheta^{0}=\frac{1}{1-\frac{\vartheta^{1}}{\vartheta^{0}}};\\
\vartheta^{1}=[\varsigma^{1}]^{2}(1-\frac{\vartheta^{1}}{\vartheta^{0}}),
\end{cases}
\end{equation}
where $\vartheta^{1}=[\frac{r_{l^{1}}}{\tau}]^{2}=[\frac{r_{v^{1}}}{r_{c}}]^{2}$ and $\vartheta^{0}=[\frac{r_{l^{0}}}{\tau}]^{2}=[\frac{r_{v^{0}}}{r_{c}}]^{2}$. The solutions are:
\begin{equation}
\begin{cases}
(\vartheta^{1}_{1},\vartheta^{0}_{1})=(\frac{-1-\sqrt{1+4[\varsigma^{1}]^{2}}}{2},\frac{1-\sqrt{1+4[\varsigma^{1}]^{2}}}{2});\\
(\vartheta^{1}_{2},\vartheta^{0}_{2})=(\frac{-1+\sqrt{1+4[\varsigma^{1}]^{2}}}{2},\frac{1+\sqrt{1+4[\varsigma^{1}]^{2}}}{2}).
\end{cases}
\end{equation}
From $(\vartheta^{1}_{2}, \vartheta^{0}_{2})$ with considering positive solutions, The Lorentz factor is then given by:
\begin{equation}
\begin{aligned}
&\gamma_{Ltz}^{2}=\frac{1}{1-\frac{\vartheta^{1}_{2}}{\vartheta^{0}_{2}}}=\frac{1}{2}(1+\sqrt{1+4[\varsigma^{1}]^{2}}).
\end{aligned}
\end{equation}

The form and structure of Eq. (30) render it particularly suitable to be interpreted as an iterative equation, as it demonstrates how a system can approximate a state through recursive calculations under specific conditions. The value of \(\vartheta_0\) depends on the preceding value of \(\vartheta_1\), and conversely, the value of \(\vartheta_1\) depends on the preceding value of \(\vartheta_0\). This interdependence creates a recursive or cyclic relationship. Through repeated iterations, one can observe whether \(\vartheta_0\) and \(\vartheta_1\) converge towards a stable value, known as a fixed point. The two solutions in Eq. (31) represent fixed points, with the positive one being a stable point, as determined by analyzing the Jacobian matrix. It provides a framework for modeling the evolution of a system over time through successive calculations. This iterative process helps in approaching a system's state under specific conditions, emphasizing the recursive relationships in spacetime dynamics and highlighting the mutual dependency between \(\vartheta_0\) and \(\vartheta_1\), which are scaling variables associated with spacetime measurements. This interdependency might illustrate the stability of these variables through iterative methods, offering potential insights into how spacetime could stabilize or reach fixed points when subjected to quantum fluctuations or scaling transformations.

The Lorentz factor can be obtained from a basic perspective. $\bar{\gamma}_{Ltz}$ is introduced to act as the Lorentz factor. Therefore, the micro space length measurement contracted by scale $\bar{\gamma}_{Ltz}$ gives $r_{l^{1}}=\sigma^{1}/\bar{\gamma}_{Ltz}=\varsigma^{1} \tau/\bar{\gamma}_{Ltz}$, while the micro time inflated by $\bar{\gamma}_{Ltz}$ gives $r_{T}=r_{l^{0}}/r_{c}=r_{t}\bar{\gamma}_{Ltz}$ (i.e. $r_{l^{0}}=r_{c}r_{T}=r_{c}r_{t}\bar{\gamma}_{Ltz}=\tau\bar{\gamma}_{Ltz}$). $v^{1}=\varsigma^{1}r_{c}/\bar{\gamma}_{Ltz}^{2}$. Since there is no contraction effect of the micro length along axis $i>1$, $r_{l}^{i>1}=\sigma^{i>1}=\varsigma^{i>1}\tau$. Using the D-dimension local flat spacetime matric $\eta_{\alpha\beta}$, we can get
\begin{small}
\begin{equation}
\kappa_{\alpha\beta}=\pm
\begin{pmatrix}
\bar{\gamma}^{2}_{Ltz} & 0 & 0 & \cdots & 0 \\
0 & -[\varsigma^{1}]^{2}/\bar{\gamma}^{2}_{Ltz} & 0 & \cdots & 0 \\
0 & 0 & -[\varsigma^{2}]^{2} & \cdots & 0 \\
\vdots & \vdots & \vdots & \ddots & \vdots\\
0 & 0 & 0 & \cdots & -[\varsigma^{D-1}]^{2}
\end{pmatrix}
.
\end{equation}
\end{small}

From the constraint of $\Sigma_{\alpha,\beta}\kappa_{\alpha\beta}=1$, we get the constrained relation connecting the dimension and components of spacetime,
\begin{equation}
\begin{aligned}
1&=\pm\{\bar{\gamma}^{2}_{Ltz}-[\varsigma^{1}]^{2}/\bar{\gamma}^{2}_{Ltz}-\sum_{k=2}^{D-1}[\varsigma^{k}]^{2}\}.
\end{aligned}
\end{equation}
We call $\kappa_{\alpha\beta}$ the spacetime structure coefficients. Eq. (34) can be deemed as an equation of $\bar{\gamma}^{2}_{Ltz}$ and its solutions are
\begin{equation}
\begin{aligned}
\bar{\gamma}^{2}_{Ltz}&=\frac{1}{2}\{(\sum_{k=2}^{D-1}[\varsigma^{k}]^{2}\pm1)\pm\sqrt{4[\varsigma^{1}]^{2}+(\sum_{k=2}^{D-1}[\varsigma^{k}]^{2}\pm1)^{2}} \}.
\end{aligned}
\end{equation}
The limit of light speed needs $\bar{\gamma}^{2}_{Ltz}>0$ in Minkowski spacetime. The physical significance of this equation lies in its role in understanding how spacetime fluctuations at microscopic scales can be related to the relativistic effects of time dilation and length contraction. The Lorentz factor is a key component in the theory of relativity, used to describe how measurements of time and space differ for observers in different inertial frames. This equation demonstrates that the Lorentz factor depends on the scaling parameters $\varsigma^{k}$ which represent the fluctuations in spacetime geometry across different dimensions. The positive value of $\bar{\gamma}^{2}_{Ltz}$ is necessary to ensure the consistency of the model with the principles of relativity, particularly in preserving the limit on the speed of light. This relationship highlights how quantum fluctuations at the Planck scale can impact classical relativistic concepts, providing fundamental insight into the nature of spacetime.

Considering two limit situations for $v^{1}=0$ and $r_{c}$. If $v^{1}=0$, it corresponds to that $\varsigma^{1}=0$ and $\bar{\gamma}_{Ltz}^{2}=\sum_{k=2}^{D-1}[\varsigma^{k}]^{2}\pm1$ or $\bar{\gamma}_{Ltz}^{2}= 0$. Consequently, $\sigma^{1}=0$, signifying there is no spatial component in this direction. Additionally, we have $\sum_{k=2}^{D-1}[\sigma^{k}]^{2}=(\bar{\gamma}_{Ltz}^{2}\mp1)\tau^{2}$ for the positive solution. If $v^{1}=r_{c}$, needing $\varsigma^{1}=\bar{\gamma}_{Ltz}^{2}$ and following with $\sigma^{1}=\bar{\gamma}_{Ltz}^{2}\tau^{2}$, then $\sum_{k=2}^{D-1}[\varsigma^{k}]^{2}=\mp1$ and $\sum_{k=2}^{D-1}[\sigma^{k}]^{2}=\mp\tau^{2}$. 

There is a generalization of Eq. (30). Based on Eq. (35), the generalized Lorentz inflation/contraction transformation becomes
\begin{equation}
\begin{cases}
&\vartheta^{0}=\frac{\sum_{k=2}^{D-1}[\varsigma^{k}]^{2}\pm1}{1-\frac{\vartheta^{1}}{\vartheta^{0}}};\\
&\vartheta^{1}=\frac{[\varsigma^{1}]^{2}}{\sum_{k=2}^{D-1}[\varsigma^{k}]^{2}\pm1}(1-\frac{\vartheta^{1}}{\vartheta^{0}}),
\end{cases}
\end{equation}
where $\vartheta^{0}\vartheta^{1}=[\varsigma^{1}]^{2}$. 

Comparing Eq. (35) to Eq. (32), when $\sum_{k=2}^{D-1}[\varsigma^{k}]^{2}$ equals 0 for time-like metric while 2 for space-like metric, we get positive $\bar{\gamma}_{Ltz}=\gamma_{Ltz}$. And then we have expressions
\begin{equation}
\begin{aligned}
&[\varsigma^{1}]^{2}=\frac{r_{c}^{2}[v^{1}]^{2}}{(r_{c}^{2}-[v^{1}]^{2})^{2}}=\frac{[\vartheta^{0}]^{2}[\vartheta^{1}]^{2}}{([\vartheta^{0}]^{2}-[\vartheta^{1}]^{2})^{2}};
\end{aligned}
\end{equation}
\begin{equation}
\begin{aligned}
&\gamma_{Ltz}^{2}-[\varsigma^{1}]^{2}\gamma_{Ltz}^{-2}=1;
\end{aligned}
\end{equation}
\begin{equation}
\begin{aligned}
&v^{1}=\frac{r_{l^{1}}}{r_{T}}=\frac{\varsigma^{1}}{\gamma_{Ltz}^{2}}r_{c}=\frac{r_{v^{1}}}{\gamma_{Ltz}}=\frac{\varsigma^{1}r_{c}}{\frac{1}{2}(1+\sqrt{1+4[\varsigma^{1}]^{2}})};
\end{aligned}
\end{equation}
\begin{equation}
\begin{aligned}
&v^{i>1}=\frac{r_{l}^{i>1}}{r_{T}}=\frac{\varsigma^{i>1}}{\gamma_{Ltz}}r_{c}=\frac{r_{v^{i>1}}}{\gamma_{Ltz}};
\end{aligned}
\end{equation}
\begin{equation}
\begin{aligned}
&v^{0}=\frac{r_{l^{0}}}{r_{T}}=r_{c}.
\end{aligned}
\end{equation}
The speed $v^{i>2}$ can be defined as the speed of fluctuation in micro-length measurements ($\sigma^{i>2}$) associated with the fluctuation on $i=1$ axis with a speed of $v^{1}$ ($0 \leq v^{1} \leq r_{c}$, TABLE 1 illustrates the relationships in the limiting situation for $v^{1}=0$ and $r_{c}$). This association is expressed via $v^{\alpha}$ being constrained by the relationship of
\begin{equation}
\begin{aligned}
&\frac{r_{c}^{2}}{\gamma_{Ltz}^{2}}=\pm([v^{0}]^{2}-\sum_{k=1}^{D-1}[v^{k}]^{2}),
\end{aligned}
\end{equation}
where $\sum_{k=2}^{D-1}[v^{k}]^{2}=\frac{2r_{c}^{2}}{\gamma_{Ltz}^{2}}=2(r_{c}^{2}-[v^{1}]^{2})$ for space-like metric and 0 for time-like metric.

\begin{table}
\caption{The relationships in the limit-situation}\label{tab:table}
\begin{small}
\begin{tabular}{cccccccc}
\hline
$v^{1}=0$  & $\gamma_{Ltz}^{2}=1$   & $\vartheta^{1}=0$ & $\varsigma^{1}=0$ & $\vartheta^{0}=1$ \\
\hline
$v^{1}=r_{c}$  & $\gamma_{Ltz}^{2}=\infty$   & $\vartheta^{1}=\vartheta^{0}$ & $\varsigma^{1}=\infty$ & $\vartheta^{0}=\infty$ \\
\hline
\end{tabular}
\end{small}
\end{table}

\section{Micro measurement in spacetime with Schwarzschild Metric}
The Schwarzschild metric describes the spacetime around a non-rotating, spherically symmetric black hole. If considering the black hole with Schwarzschild Metric \cite{bibitem14}, the space-like line element has the form of
\begin{equation}
\begin{aligned}
d\lambda^{2}=-(1-\frac{r_{s}}{r})c^{2}dt^{2}+(1-\frac{r_{s}}{r})^{-1}dr^{2}+r^{2}d\Omega^{2}
\end{aligned}
\end{equation}
in Schwarzschild coordinates, where $d\Omega^{2}=d\theta^{2}+sin^{2}(\theta)d\phi^{2}$. And $r_{s}=\frac{2GM}{c^{2}}$ is the Schwarzschild radius at where the event horizon is. $G$ is the gravitational constant, $M$ is the object mass, and $c$ is the speed of light. In the presence of gravitational fields, measurements of space and time are affected differently compared to special relativity. Specifically, gravitational fields can cause contraction effects in spatial measurements and dilation effects in time measurements. These effects arise due to the curvature of spacetime in general relativity. In regions with strong gravity, time appears to slow down (gravitational time dilation), while space can be warped, affecting perceived distances. These phenomena differ from the effects predicted by special relativity, which deals with time dilation and length contraction due to relative motion. However, in the absence of gravitational fields or in regions with weak gravity, the predictions of general relativity converge to those of special relativity. In such scenarios, spacetime is essentially flat, and the familiar Lorentz transformations apply, with the Lorentz factor $\bar{\gamma}_{Ltz}$ describing time dilation and length contraction.

Define a new factor $\gamma_{g}$ to replace $\bar{\gamma}_{Ltz}$ to act as the Lorentz factor in the spacetime of Schwarzschild Metric to measure the micro length contraction and micro time inflation effect. It can be seen as a correction factor with respect to $\bar{\gamma}_{Ltz}$ when measuring in a gravitational field. Using the constraint of $\Sigma_{\alpha,\beta}\kappa_{\alpha\beta}=1$, we get
\begin{equation}
\begin{aligned}
\gamma_{g}^{2}=(1-\frac{r_{s}}{r})^{-1}\bar{\gamma}^{2}_{Ltz}.
\end{aligned}
\end{equation}
When $M=0$ or $r\rightarrow\infty$, $\gamma^{2}_{g}=\bar{\gamma}^{2}_{Ltz}$. When $r=r_{s}$, $\gamma^{2}_{g}=\infty$. When $r=0$, $\gamma^{2}_{g}=0$. $\gamma^{2}_{g}$ is remaining positive when $\bar{\gamma}^{2}_{Ltz}<0$ at $r<r_{s}$ and $\bar{\gamma}^{2}_{Ltz}>0$ at $r>r_{s}$. It can be seen that $\gamma_{g}$ describes a dual effect of gravitational field in the general relativity frame and local motion in the special relativity frame. From this generalized Lorentz factor $\gamma_{g}$ in Schwarzschild Metric, we can still keep Eq. (34) remaining. Consequently, the physical laws satisfy the Lorentz transformation when the physical variables are defined to be scaled by $\gamma_{g}$.

There is also a generalization of Lorentz inflation/contraction transformation in Schwarzschild Metric. Namely,
\begin{equation}
\begin{cases}
&\vartheta^{0}=(1-\frac{r_{s}}{r})^{-1}\frac{\sum_{k=2}^{D-1}[\varsigma^{k}]^{2}\pm1}{1-\frac{\vartheta^{1}}{\vartheta^{0}}};\\
&\vartheta^{1}=(1-\frac{r_{s}}{r})^{-1}\frac{[\varsigma^{1}]^{2}}{\sum_{k=2}^{D-1}[\varsigma^{k}]^{2}\pm1}(1-\frac{\vartheta^{1}}{\vartheta^{0}}),
\end{cases}
\end{equation}
where $\vartheta^{0}\vartheta^{1}=[\varsigma^{1}]^{2}(1-\frac{r_{s}}{r})^{-2}$. 

The mass of a particle moving along $r$ scaled by $\gamma_{g}$ is expressed by
\begin{equation}
\begin{aligned}
&m^{2}_{g}=\gamma_{g}^{2}m^{2}_{0}=(1-\frac{r_{s}}{r})^{-1}\bar{\gamma}^{2}_{Ltz}m^{2}_{0}=\bar{\gamma}^{2}_{Ltz}m^{2}_{g0},
\end{aligned}
\end{equation}
where $m^{2}_{g0}=(1-\frac{r_{s}}{r})^{-1}m^{2}_{0}$ is introduced as the measured mass of the particle at instantaneous rest in a gravitational field with Schwarzschild metric and $m_{0}$ is the mass of the particle at rest in a flat metric. When $r\rightarrow\infty$ or $M=0$, $m^{2}_{g0}=m^{2}_{0}$. When $r\rightarrow0$, $m^{2}_{g0}=0$. Therefore, the energy-momentum relationship measured locally in the gravitational field with the Schwarzschild Metric becomes
\begin{equation}
\begin{aligned}
&E^{2}_{g}=p^{2}_{g}c^{2}+m^{2}_{g0}c^{4},
\end{aligned}
\end{equation}
where $E^{2}_{g}=\bar{\gamma}^{2}_{Ltz}(1-\frac{r_{s}}{r})^{-1}m^{2}_{0}c^{4}=\bar{\gamma}^{2}_{Ltz}m^{2}_{g0}c^{4}$. And $p^{2}_{g,1}=\bar{\gamma}^{2}_{Ltz}(1-\frac{r_{s}}{r})^{-1}m^{2}_{0}v^{2}_{g1}=\bar{\gamma}^{2}_{Ltz}m^{2}_{g0}v^{2}_{g,1}$ with $v_{g,1}=\frac{\varsigma^{1}r_{c}}{\gamma_{g}^{2}}$ and $p^{2}_{g,k>1}=\bar{\gamma}^{2}_{Ltz}m^{2}_{g0}v^{2}_{g,k>1}$ with $v_{g,k>1}=\frac{\varsigma^{k>1}r_{c}}{\gamma_{g}}$. Thus, $p^{2}_{g}=\sum_{k=1}^{D-1} p^{2}_{g,k}$. This energy-momentum relationship scaled by $(1-\frac{r_{s}}{r})$ on both sides can be back to the one in Minkowski spacetime. When $r\rightarrow\infty$ or $M=0$, the former is also back to the latter. When $r>>r_{s}$, we get
\begin{equation}
\begin{aligned}
&E^{2}_{g}=p^{2}_{g}c^{2}+m^{2}_{0}c^{4}[1+\frac{r_{s}}{r}+(\frac{r_{s}}{r})^{2}+O(3)],
\end{aligned}
\end{equation}

\section{Scaling Geodesic Equations and Einstein Field Equations}
Geodesic equations \cite{bibitem8,bibitem9, bibitem10, Tsamparlis, Hackmann} are fundamental in the study of general relativity and differential geometry. They describe the path that an object follows under the influence of gravity alone, without any other forces acting on it. They are crucial for understanding the motion of objects in a gravitational field, serving as the foundation for analyzing how gravity shapes the trajectories of particles and light. They provide insights into both macroscopic phenomena, like the orbits of planets, and microscopic effects, such as the influence of gravity on quantum particles. As such, geodesics are integral to bridging the gap between classical and quantum descriptions of gravity.

The geodesic equations can be written in scaling form on a micro-scale via operators' transformations. Firstly, by using Eq. (8), the Christoffel symbols can be represented in the form of
\begin{equation}
\begin{aligned}
\mathbf{\Gamma}^{\alpha}_{\beta\gamma}&=\frac{1}{2}g^{\alpha\mu}\frac{1}{dx_{0}}(\hat{a}_{\beta}\frac{\partial g_{\gamma\mu}}{\partial X^{\beta}}+\hat{a}_{\gamma}\frac{\partial g_{\beta\mu}}{\partial X^{\gamma}}+\hat{a}_{\mu}\frac{\partial g_{\beta\gamma}}{\partial X^{\mu}})\\
&=\frac{1}{dx_{0}}\Gamma^{\alpha}_{\beta\gamma}
\end{aligned}
\end{equation}
with Einstein notation, where $\Gamma^{\alpha}_{\beta\gamma}$ are defined as dimensionless scaling Christoffel symbols. Combining the Eq. (3) and Eq. (4), the scaling geodesic equations are obtained by
\begin{equation}
\begin{aligned}
(\bar{L}^{\alpha}-L^{\alpha})+\Gamma^{\alpha}_{\beta\gamma}L^{\alpha}L^{\beta}=0
\end{aligned}
\end{equation}
with Einstein notation. The equation demonstrates how local geodesic paths (which describe the shortest path between points in spacetime) evolve with scaling functions, offering insights into the complex structure of spacetime at microscopic scales. 

In the framework of general relativity, gravity is conceptualized not as a force but as an outcome of the curvature of spacetime, with the stress-energy tensor serving as the source of this curvature. Within the geometric interpretation of gravity, geodesics are indicative of the curved geometry inherent in spacetime. All length-minimizing curves are geodesics, and all geodesics are length-minimizing, at least locally \cite{bibitem15}. The local geodesic equations can be viewed as a constraint on microscopic spacetime measurements at a local level. When representing the measurement of spacetime microelements through scaling functions, the geodesic equations represented by scales elucidate the restrictions on nontrivial fluctuations of scales. 

The scaled Einstein Field Equations represented by scales $X^{\alpha}$ can be obtained in the same way. Which characterizes how the scales of micro geometric spacetime measurements are influenced by stress-energy tensor.

By considering a transformation of a linear mapping between coordinate $x^{\alpha}$ and scale $\bar{X}^{\alpha}$. The latter are new scales obtained by rescaling $\zeta^{\alpha}$ on $X^{\alpha}$. $\zeta^{\alpha}$ are the rescaling factors on differential operators, establishing a connection between coordinate $x^{\alpha}$ and scale $\bar{X}^{\alpha}$ and making the micro measurements being carried out in a linear way, i.e., 
\begin{equation}
\begin{aligned}
&dx^{\alpha}=d\bar{X}^{\alpha}dx^{\alpha}_{0}.
\end{aligned}
\end{equation}
where $\bar{X}^{\alpha}$ denotes a scaled version of coordinate $x^{\alpha}$ in fact of our conventional understanding of the definition of measurement based on the unit system. We call these mapping conditions on rescaling factors $\zeta^{\alpha}$. This will bring the spacetime structure measured by nonlinear kernel $L^{\alpha}(X^{\alpha})$ back to the linear kernel $\bar{X}^{\alpha}$. The spacetime fluctuation determined by the latter can equivalently represent the erratic point ($x^{\alpha}$) in spacetime. This greatly simplifies the complexity of handling the problems at the micro-scale and facilitates the extraction of intrinsic properties of the dynamic systems involved.

It is essential to emphasize that $x^{\alpha}$ represents the position of a point in the coordinate frame on spacetime. However, $\bar{X}^{\alpha}$ and $X^{\alpha}$ denote the measuring scales of the micro-interval of the fluctuating spacetime near a point along the $\alpha$-axis direction. Through a linear mapping between $x^{\alpha}$ and $\bar{X}^{\alpha}$, the latter characterizes linear scaling measurements in the spacetime defined by the former. On the other hand, $X^{\alpha}$ describes a spacetime with a structure of nonlinear measurement. 

Comparing Eq. (8) and Eq. (13), the mapping conditions
\begin{equation}
\begin{aligned}
&\frac{\partial}{\partial x^{\alpha}}=\frac{1}{dx_{0}^{\alpha}}\frac{\partial}{\partial \bar{X}^{\alpha}}
\end{aligned}
\end{equation}
could be satisfied, resulting in a transformation making the microelement measurements from a nonlinear mode into a linear one. These linear conditions can be equivalently expressed as $\frac{1}{\zeta^{\alpha}}=\hat{a}_{\alpha}$ and then we get equations of
\begin{equation}
\begin{aligned}
&\frac{dL^{\alpha}}{dX^{\alpha}}=\zeta^{\alpha}(\frac{\bar{L}^{\alpha}}{L^{\alpha}}-1).
\end{aligned}
\end{equation}
When $\frac{dL^{\alpha}}{dX^{\alpha}}=0$, then corresponds to \(\zeta^\alpha=1\), the spacetime nature is back to the classical non-fluctuation (stable) structure. 

These equations can be regarded as nonlinear feedback equations, where the differences in the state of $L^{\alpha}$ at different scaling scales drive the dynamic evolution of the system, leading to complex behaviors such as fluctuations, chaos, or bifurcations, depending on the system parameters. Rescaling factor \( \zeta^\alpha \) modulates the system's sensitivity, affecting how the output \( \bar{L}^\alpha \) responds to inputs at the original scale \( X^\alpha \), affecting the stability and dynamic behavior of the system. Similar structures of Eq. (53) can be found in delay differential equations \cite{Richard}, chaotic systems\cite{Lathrop}, and proportional feedback models\cite{Astrom}. These types of equations are commonly used to describe adaptive systems, nonlinear control systems, and dynamic systems with complex feedback structures. The form of the nonlinear feedback equations suggests the existence of adaptive and dynamic behaviors in spacetime geometry, prompting us to explore how more complex feedback structures manifest the evolution of spacetime at a micro-scale.

When $L^{\alpha}=[X^{\alpha}]^{q^{\alpha}}$ with $q^{\alpha}$ being exponents that characterize a self-similarity and scaling behavior of spacetime fluctuations along the $\alpha$-axis. In this situation, Eq. (53) becomes
\begin{equation}
\begin{aligned}
&[\zeta^{\alpha}]^{q^{\alpha}+1}-\zeta^{\alpha}-q^{\alpha}[X^{\alpha}]^{q^{\alpha}-1}=0.
\end{aligned}
\end{equation}
When $q^{\alpha}=1$, we get $\zeta^{\alpha}=\varphi=\frac{1\pm\sqrt{5}}{2}\approx1.618034(-0.618034)$, it is the golden ratio. The nonlinear form of this equation highlights the complexity of scaling effects, emphasizing the intricate relationship between scaling factors and exponents.

The subsequent instances are the scaled geodesic equations and scaled Einstein field equations within the linear measurement spacetime, as well as the constraint equations derived from their linear mapping conditions.

With the linear mapping conditions of
\begin{equation}
\begin{cases}
&\frac{\partial}{\partial x^{\alpha}}=\frac{1}{dx_{0}^{\alpha}}\frac{\partial}{\partial \bar{X}^{\alpha}},\\
&\frac{dx^{\alpha}}{d\lambda}=\frac{d\bar{X}^{\alpha}}{d\bar{X}},\\
&\frac{d^{2}x^{\alpha}}{d\lambda^{2}}=\frac{d^{2}\bar{X}^{\alpha}}{d\bar{X}^{2}}\frac{1}{d\lambda_{0}},
\end{cases}
\end{equation}
the geodesic equations can be derived in the equivalent scaling form, e.i.
\begin{equation}
\begin{aligned}
\frac{d^{2}\bar{X}^{\alpha}}{d\bar{X}^{2}}+\bar{\Gamma}^{\alpha}_{\beta\gamma}\frac{d\bar{X}^{\beta}}{d\bar{X}}\frac{d\bar{X}^{\gamma}}{d\bar{X}}=0
\end{aligned}
\end{equation}
with Einstein notation, where scaling Christoffel symbols are
\begin{equation}
\begin{aligned}
\bar{\Gamma}^{\alpha}_{\beta\gamma}=\frac{1}{2}\bar{g}^{\alpha\mu}(\frac{\partial \bar{g}_{\gamma\mu}}{\partial \bar{X}^{\beta}}+\frac{\partial \bar{g}_{\beta\mu}}{\partial \bar{X}^{\gamma}}+\frac{\partial \bar{g}_{\beta\gamma}}{\partial \bar{X}^{\mu}})
\end{aligned}
\end{equation}
with 
\begin{equation}
\begin{aligned}
\bar{g}_{\alpha\beta}(\bar{X}^{\mu})=g_{\alpha\beta}(x^{\mu}\rightarrow\bar{X}^{\mu}).
\end{aligned}
\end{equation}
In Eq. (55), $\frac{d\bar{X}^{\alpha}}{d\bar{X}}=\frac{d\bar{X}^{\alpha}}{dX^{\alpha}}\frac{dX^{\alpha}}{dX}\frac{dX}{d\bar{X}}=(\zeta^{\alpha}+X^{\alpha}\frac{d\zeta^{\alpha}}{dX^{\alpha}})\frac{dX^{\alpha}}{dX}\frac{1}{\chi}$. Utilizing the definitions of $r_{l}^{\alpha}$ and $r_{l}$ and the expression of $\frac{dr_{l^{\alpha}}}{dr_{l}}$, we get $\frac{dX^{\alpha}}{dX}=\frac{\bar{L}^{\alpha}-L^{\alpha}}{\bar{\tau}-\tau}\frac{d\tau/dX}{dL^{\alpha}/dX^{\alpha}}$. Thus, the linearized requirements of $\frac{dx^{\alpha}}{d\lambda}=\frac{d\bar{X}^{\alpha}}{d\bar{X}}$ and $\frac{1}{\zeta^{\alpha}}=\hat{a}_{\alpha}$ result in
\begin{equation}
\begin{aligned}
&1+\frac{X^{\alpha}}{\zeta^{\alpha}}\frac{d\zeta^{\alpha}}{dX^{\alpha}}=\chi\hat{a},
\end{aligned}
\end{equation}
The requirement of $\frac{d^{2}x^{\alpha}}{d\lambda^{2}}=\frac{d^{2}\bar{X}^{\alpha}}{d\bar{X}^{2}}\frac{1}{d\lambda_{0}}=\frac{1}{d\lambda_{0}}\frac{d}{dX}(\frac{d\bar{X}^{\alpha}}{d\bar{X}})\frac{dX}{d\bar{X}}$ further leads to
\begin{equation}
\begin{aligned}
&\frac{d\tau}{dX}=(\frac{1}{\hat{a}}-\chi)(\frac{\bar{L}^{\alpha}}{L^{\alpha}}-1),
\end{aligned}
\end{equation}
If linear mapping condition is used for proper measurement, $\hat{a}=\frac{1}{\chi}$, resulting $\zeta^{\alpha}$ and $\tau$ are constants. In addition, from $\frac{1}{\zeta^{\alpha}}=\hat{a}_{\alpha}$, we can obtain Eq. (53), with $\zeta^{\alpha}$ independent on $X^{\alpha}$ once $\hat{a}=\frac{1}{\chi}$.

The geodesic equations formulated in terms of coordinates \( x^{\alpha} \) describe a continuous and stable spacetime. In contrast, the scaling geodesic equations Eq. (50) based on scales \( X^{\alpha} \) characterize a fluctuating spacetime at the microscopic level. Furthermore, the scaling geodesic equations Eq. (56) defined with scales \( \bar{X}^{\alpha} \) illustrate a fluctuating spacetime at the micro-scale with the linear mapping conditions.  

With the linear mapping conditions of
\begin{equation}
\begin{cases}
&\frac{\partial}{\partial x^{\alpha}}=\frac{1}{dx_{0}^{\alpha}}\frac{\partial}{\partial \bar{X}^{\alpha}},\\
&\frac{\partial^{2} }{\partial x^{\alpha}\partial x^{\beta}}=\frac{1}{dx_{0}^{\alpha}dx_{0}^{\beta}}\frac{\partial^{2}}{\partial \bar{X}^{\alpha}\partial \bar{X}^{\beta}},
\end{cases}
\end{equation}
using $\frac{1}{\zeta^{\beta}}=\hat{a}_{\beta}$ and $\frac{1}{[\zeta^{\alpha}]^{2}}\frac{d\zeta^{\alpha}}{dX^{\alpha}}=\hat{b}_{\alpha}$, under the transformation of $\zeta^{\alpha}$ for differential operators between $x^{\alpha}$ and $\bar{X}^{\alpha}$, we can similarly get the scaling Einstein field equations in the absence of any matter,
\begin{equation}
\begin{aligned}
\bar{G}_{\mu\nu}=\bar{R}_{\mu\nu}-\frac{1}{2}\bar{g}_{\mu\nu}\bar{R}=0,
\end{aligned}
\end{equation}
and the scaling vacuum Einstein equations in the presence of a cosmological constant $\bar{\Lambda}$,
\begin{equation}
\begin{aligned}
\bar{G}_{\mu\nu}=\bar{R}_{\mu\nu}-\frac{1}{2}\bar{g}_{\mu\nu}\bar{R}+\bar{\Lambda}\bar{g}_{\mu\nu}=0,
\end{aligned}
\end{equation}
where scaling Riemann curvature tensor
\begin{equation}
\begin{aligned}
\bar{R}^{\rho}_{\sigma\mu\nu}=\frac{\partial}{\partial\bar{X}^{\mu}}\bar{\Gamma}^{\rho}_{\nu\sigma}-\frac{\partial}{\partial\bar{X}^{\nu}}\bar{\Gamma}^{\rho}_{\mu\sigma}+\bar{\Gamma}^{\rho}_{\mu\lambda}\bar{\Gamma}^{\lambda}_{\nu\sigma}
+\bar{\Gamma}^{\rho}_{\nu\lambda}\bar{\Gamma}^{\lambda}_{\mu\sigma}
\end{aligned}
\end{equation}
and $\bar{R}=\bar{g}^{\mu\nu}\bar{R}_{\mu\nu}$, with Einstein notation. Therefore, we conclude that the scaling Einstein field equations in the absence of any matter and in the presence of a cosmological constant $\bar{\Lambda}$ require $\zeta^{\alpha}$ to satisfy
\begin{equation}
\begin{cases}
&\frac{dL^{\alpha}}{dX^{\alpha}}=\zeta^{\alpha}(\frac{\bar{L}^{\alpha}}{L^{\alpha}}-1);\\
&\frac{d\zeta^{\alpha}}{dX^{\alpha}}=[\zeta^{\alpha}]^{2}\hat{b}_{\alpha}.
\end{cases}
\end{equation}

For the Einstein field equations with some specific stress-energy tensor,  more conditions embedded in the stress-energy tensor need to be considered in the linear mapping conditions.

\section{The golden ratio at the micro-scale}
The micro distance measurement with $L^{\alpha}=X^{\alpha}$ corresponds to a linear mode, expressed as $dx^{\alpha}=X^{\alpha}dx^{\alpha}_{0}$. By adding the linear mapping $dx^{\alpha}=d\bar{X}^{\alpha}dx^{\alpha}_{0}$, we obtain the following equations:
\begin{equation}
\begin{cases}
&x^{\alpha}=X^{\alpha}dx^{\alpha}_{0}+x^{\alpha}_{0};\\
&x^{\alpha}=\bar{X}^{\alpha}dx^{\alpha}_{0}+\bar{x}^{\alpha}_{0},
\end{cases}
\end{equation}
where $dx^{\alpha}=x^{\alpha}-x^{\alpha}_{0}$ and $x^{\alpha}$ are the position coordinates relative to reference position coordinates  $x^{\alpha}_{0}$ or $\bar{x}^{\alpha}_{0}$, By letting $dx^{\alpha}_{0}=x^{\alpha}_{0}-\bar{x}^{\alpha}_{0}$, we get
\begin{equation}
\begin{aligned}
&[(\bar{X}^{\alpha}-X^{\alpha})-1]dx^{\alpha}_{0}=[(\varphi-1)X^{\alpha}-1]dx^{\alpha}_{0}=0,
\end{aligned}
\end{equation}
resulting in $\frac{dx^{\alpha}}{dx^{\alpha}_{0}}=X^{\alpha}=\frac{1}{\varphi-1}=\varphi$. Thus $\bar{X}^{\alpha}=\zeta^{\alpha} X^{\alpha}=\varphi X^{\alpha}=\varphi^{2}$. Introduce $d\bar{x}^{\alpha}=x^{\alpha}-\bar{x}^{\alpha}_{0}$, we get $\frac{d\bar{x}^{\alpha}}{dx^{\alpha}}=\frac{\bar{X}^{\alpha}}{X^{\alpha}}=\varphi$. This is illustrated in FIG. 1, where \( \varphi \) is the golden ratio, indicating that within the spacetime of linear scale measurement, the golden ratio naturally arises at the microscopic scale. 

To illustrate the linear measurement of micro distance, consider selecting two points $\bar{x}_{0}$ and $x_{0}$ that are sufficiently close in the coordinate space, the interval $dx_{0}=x_{0}-\bar{x}_{0}$ between these two points is defined as the reference length. When a third point is selected, two microscopic intervals, $d\bar{x}$ and $dx$ can be defined. The ratio of these intervals is the golden ratio. We propose that micro-distance measurements in a defined 1-dimensional coordinate system can occur in two distinct scenarios: one where the third point is selected outside the reference interval and another where it is chosen within the interval. These scenarios correspond to the positive $\varphi_{R}=\frac{1+\sqrt{5}}{2}$ and negative $\varphi_{L}=\frac{1-\sqrt{5}}{2}$ solutions of the equation Eq. (53) with $L^{\alpha}=X^{\alpha}$, as shown in FIG. 1 (a) and (b) respectively. Here, the subscript "R" indicates that $x$ is to the right of the reference point $x_{0}$, while "L" is to the left. 

\begin{figure}[!htb]
    \includegraphics
    [width=1.0\hsize]
      {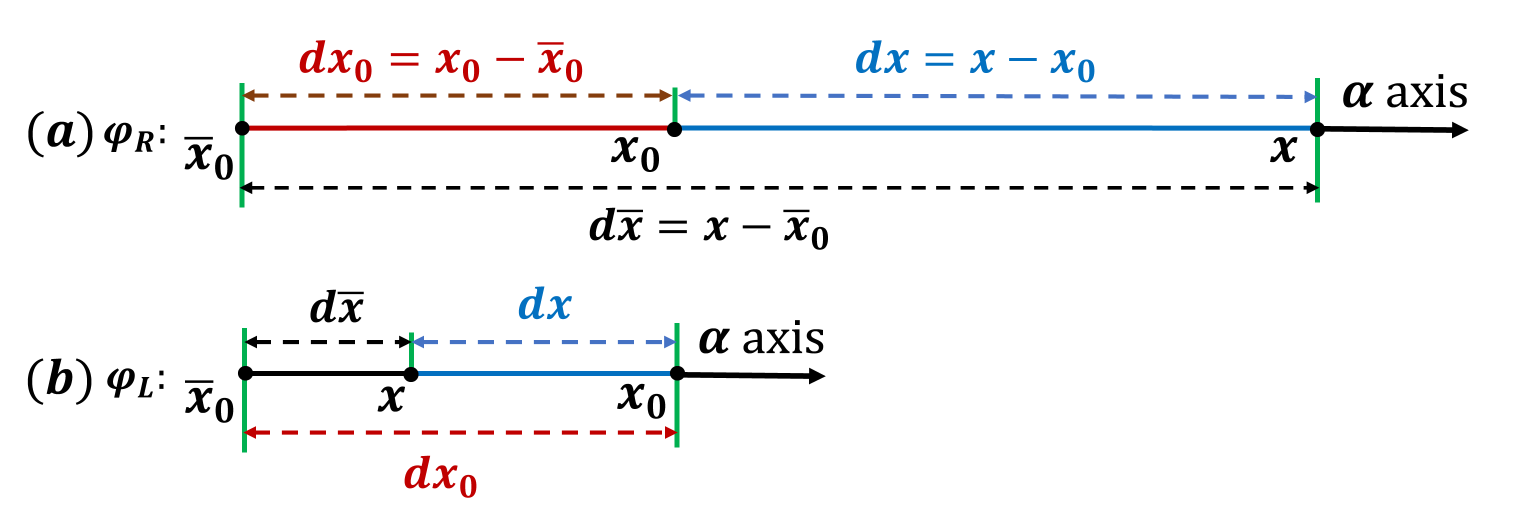}
    \caption{Golden section at the microscopic scale in the spacetime of linear scale measurement. } 
    \label{fig:Block_Schematic}
\end{figure}

In this framework, for external measurements, $dx^{\alpha}_{0}$ functions as the minimum reference length, limiting the measurement of micro-lengths smaller than $dx^{\alpha}_{0}$. Conversely, for internal measurements, $dx^{\alpha}_{0}$ functions as the maximum reference length, constraining the measurement of micro-lengths larger than $dx^{\alpha}_{0}$. This insight provides a clue for understanding the existence of the Planck length, which manifests as a lower limit for external measurements and an upper limit for internal measurements in the depiction of microscopic phenomena. 

In general, arbitrary \( L^{\alpha} = L^{\alpha}(X^{\alpha}) \) satisfying the linear mapping conditions are appropriate to the above analysis, making 
\begin{equation}
\begin{aligned}
&\bar{X}^{\alpha}=\zeta^{\alpha}X^{\alpha}=1+L^{\alpha}.
\end{aligned}
\end{equation}
If \( L^{\alpha} > 0 \) for the right measuring, while \( L^{\alpha} < 0 \) for the left measuring. Therefore, the metric tensor (Eq. 58) becomes $\bar{g}_{\alpha\beta}(\bar{X}^{\sigma})=g_{\alpha\beta}[x^{\sigma}\rightarrow x^{\sigma}_{0}+(\bar{X}^{\sigma}-1)dx^{\sigma}_{0}]$. The case of \( \bar{X}^{\alpha} > 1 \) (external measurements) indicates that the UV problem is avoidable for the evolution of the equations (Eq. 56, 62, 63) in spacetime described by \( \bar{X}^{\sigma} \). 

Once Eq. (53) is multiplied by $X^{\alpha}$, we get renormalization group equations \cite{Stueckelberg} for micro measurements of the micro spacetime fluctuations, namely
\begin{equation}
\begin{aligned}
&X^{\alpha}\frac{dL^{\alpha}}{dX^{\alpha}}=\beta(L^{\alpha})=(1+L^{\alpha})(\frac{\bar{L}^{\alpha}}{L^{\alpha}}-1),
\end{aligned}
\end{equation}
where $\beta(L^{\alpha})$ is the beta function and  $\beta(L^{\alpha})=0$ defines the fixed points of the measurements. At the fixed points $\bar{L}^{\alpha}=L^{\alpha}$, the beta function $\beta(L^{\alpha})$ vanishes, indicating that the fluctuations have reached equilibrium and a symmetric state, scale invariance. Deviations from this point drive the scaling behavior of the fluctuations, causing it to adjust towards the fixed point. The case $L^{\alpha} = -1$ represents a potential unique solution. Based on stability analysis, $L^{\alpha} = -1$ are saddle points ($\frac{d\beta}{dL^{\alpha}}$=0) while $\bar{L}^{\alpha} = L^{\alpha}$ are stable points ($\frac{d\beta}{dL^{\alpha}}<0$) if $L^{\alpha}\geq0$ or $<-1$. If $-1<L^{\alpha}<0$, $\bar{L}^{\alpha} = L^{\alpha}$ are unstable points. Consequently, the measurements are unstable within the region between $\bar{x}^{\alpha}_{0}$ and $x^{\alpha}_{0}$ but stable outside.
Here, $\bar{L}^{\alpha}=L^{\alpha}(\zeta^{\alpha}X^{\alpha})=L^{\alpha}(X^{\alpha})$ means scale invariant.

When \( L^{\sigma} = 1 \), it follows \( \bar{X}^{\sigma} = 2 \), resulting in \( x^{\sigma} = x^{\sigma}_{0} + dx^{\sigma}_{0} \). When \( L^{\sigma} = -1 \), \( \bar{X}^{\sigma} = 0 \), resulting in \( x^{\sigma} = x^{\sigma}_{0} - dx^{\sigma}_{0}= \bar{x}^{\sigma}_{0} \). In this case, spacetime measurements are performed in uniform and equidistant micro-intervals without fluctuations, and the physics is defined on equidistantly discrete points. 

When considering a not-allowed situation in the current framework where \( L^{\alpha}\rightarrow0 \), we have \( x^{\sigma} = x^{\sigma}_{0} \), which leads to \( g_{\alpha\beta} \rightarrow \infty \) if Eq. (25) consistently used, indicating divergence, this represents a scenario of a singularity without fluctuation. Consequently, classical continuous spacetime allows for the definition of arbitrarily small intervals (i.e., $dx^{\alpha}=0$), indicating that the definition of continuous spacetime results in singularities of spacetime measurements appearing everywhere when making arbitrarily infinitesimal measurements. 

The emergence of the golden ratio in linear micro measurements suggests that spacetime possesses an intrinsic order characterized by mapping constraints. Physics is defined at two fixed points, \( \varphi_{R} \) and \( \varphi_{L} \), with bilateral asymmetry around \( x^{\sigma}_{0} \). Generally, the mapping constraints can influence spacetime patterns or structures, making physics defined on discrete points (might be multiple) once the forms of $L^{\alpha}(X^{\alpha})$ measurements are given, since there are equations for solving $X^{\alpha}$ based on Eq. 53 and Eq. 68. There is a condition to determine the forms of \( L^{\alpha}(X^{\alpha})=\zeta^{\alpha}X^{\alpha}-1 \): \( \zeta^{\alpha} \) are constants via analyzing Eq. (59) if considering the linear mapping constraint of \( \hat{a} = \frac{1}{\chi} \) for proper length measurement, resulting in $X^{\alpha}=\frac{1}{2\zeta^{\alpha}-[\zeta^{\alpha}]^{2}}$. When solving the forms of \( L^{\alpha}(X^{\alpha}) \),  coupling with \( \zeta^{\alpha}(X^{\alpha}) \), using equations Eq. (53) and Eq. (68) without considering the constraint \( \chi\hat{a} = 1 \), it leads to the conclusion that micro measurements are performed through fluctuating functions \( L^{\alpha}(X^{\alpha}) \), indicating that physics is defined on a fluctuating spacetime.

What can be argued is that \( dx^{\alpha}_{0} \) as a critical reference length for both external and internal measurements plays a key role in addressing the problem of UV divergence \cite{bibitem1, bibitem2, Green1987, Schwartz, Weinberg, RovelliQG}. For external measurements, \( dx^{\alpha}_{0} \) acts as the minimum reference length, preventing the measurement of micro-lengths smaller than this threshold. This limitation imposes a natural cutoff on the smallest measurable scales, thus setting a lower bound for the length scale of the system. Since UV divergences in quantum field theory arise from summing over arbitrarily small distances (or high energy scales), imposing a minimum length scale like \( dx^{\alpha}_{0} \) effectively circumvents such divergences. Specifically, the Planck length, \( l_p = \sqrt{\frac{\hbar G}{c^3}} \approx 1.616 \times 10^{-35}\) meters, serves as this fundamental minimum length, providing a physical cutoff at the Planck scale and thereby avoiding UV divergences.
Conversely, for internal measurements, \( dx^{\alpha}_{0} \) functions as the maximum reference length, constraining the measurement of micro-lengths larger than this limit. This creates an upper bound for internal measurements. By setting \( dx^{\alpha}_{0} \) as the lower bound for external measurements, the framework naturally introduces a constraint on measurable length scales. This constraint inherently avoids UV divergence by preventing the system from probing arbitrarily small lengths or high energy scales, which are typically responsible for such divergences in field theories. This mechanism is akin to the non-perturbative quantum gravity frameworks like LQG, where spacetime is quantized, thereby preventing classical singularities and divergences. Moreover, it parallels string theory, wherein the finite size of strings acts as a natural cutoff for high-energy interactions, effectively avoiding UV divergence.
In summary, by incorporating \( dx^{\alpha}_{0} \) as a critical scale, this framework provides a natural means of avoiding UV divergence through physical limits on the smallest measurable lengths, resonating with the Planck scale as a fundamental boundary in quantum gravity and quantum field theories.

However, the physical significance of the mapping constraints and their connection to fundamental constants like the Planck length would require further investigation and validation within the broader framework of theoretical physics.

\section{Scaling Klein-Gordon Equation and Dirac Equation}
The Klein-Gordon and Dirac equations are essential in quantum mechanics and quantum field theory, describing the behavior of particles and fields while incorporating quantum and relativistic effects \cite{Schwartz,Weinberg}. Investigating scaling within these equations may offer insights into how particle fields behave in spacetimes with complex geometries. At extremely small scales, such as the Planck length, the classical notion of continuous spacetime might become inadequate, potentially giving rise to a network of quantized geometries \cite{RovelliQG}. If spacetime indeed has a fluctuation structure, this could lead to modifications in the behavior of fields and particles.

From the operator transformations introduced in the second section, the Klein-Gordon equation and Dirac equation respective to the scale variables at the local micro-scale can be deduced and their forms in linear measuring spacetime can be obtained with mapping constraints.
The Klein-Gordon equation \cite{bibitem16, bibitem17}
\begin{equation}
\begin{aligned}
&\{\Box-[\frac{m  c}{\hbar}]^{2}\}\phi=0
\end{aligned}
\end{equation}
can be written in scaling form of
\begin{equation}
\begin{aligned}
&\{\eta^{\mu\nu}\hat{a}_{\mu}\hat{a}_{\nu}(\frac{\partial^{2}}{\partial X^{\mu}\partial X^{\nu}}-\hat{b}_{\nu}\frac{\partial X^{\nu}}{\partial X^{\mu}}\frac{\partial}{\partial X^{\nu}})\\
&-[\frac{r_{m} r_{c}}{r_{\hbar}}]^{2}\frac{[dx_{0}]^{2}[m_{0}]^{2}[[dx_{0}]/[dt_{0}]]^{2}}{[\hbar_{0}]^{2}}\}\phi=0.
\end{aligned}
\end{equation}
Where $\Box=\eta^{\mu\nu}\frac{\partial^{2}}{\partial x^{\mu}\partial x^{\nu}}$, $\phi=\phi(x^{\alpha})$, $r_{m}=\frac{m}{m_{0}}$ being the scaling measurement for particle mass, and $r_{\hbar}=\frac{\hbar}{\hbar_{0}}$ being the scaling measurement for Planck constant. After dimensional analysis, we get
\begin{footnotesize}
\begin{equation}
\begin{aligned}
&\{\eta^{\mu\nu}\hat{a}_{\mu}\hat{a}_{\nu}(\frac{\partial^{2}}{\partial X^{\mu}\partial X^{\nu}}-\hat{b}_{\nu}\frac{\partial X^{\nu}}{\partial X^{\mu}}\frac{\partial}{\partial X^{\nu}})-[\frac{r_{m} r_{c}}{r_{\hbar}}]^{2}\}\phi=0£¬
\end{aligned}
\end{equation}
\end{footnotesize}
Here, $\phi=\phi(x^{\sigma}\rightarrow x^{\sigma}_{0}+L^{\sigma}(X^{\sigma})dx^{\sigma}_{0})$, representing the fluctuation of the $\phi$ field near $x^{\sigma}_{0}$ as driven by $L^{\sigma}$ nonlinear fluctuates.

With the linear mapping conditions of
\begin{equation}
\begin{cases}
&\frac{\partial}{\partial x^{\alpha}}=\frac{1}{dx_{0}^{\alpha}}\frac{\partial}{\partial \bar{X}^{\alpha}},\\
&\frac{\partial^{2} }{\partial x^{\alpha}\partial x^{\beta}}=\frac{1}{dx_{0}^{\alpha}dx_{0}^{\beta}}\frac{\partial^{2}}{\partial \bar{X}^{\alpha}\partial \bar{X}^{\beta}},
\end{cases}
\end{equation}
we can similarly get the scaling Klein-Gordon equation expressed as
\begin{equation}
\begin{aligned}
&\{\eta^{\mu\nu}\frac{\partial^{2}}{\partial \bar{X}^{\mu}\partial \bar{X}^{\nu}}-[\frac{r_{m} r_{c}}{r_{\hbar}}]^{2}\}\bar{\psi}_{KG}=0,
\end{aligned}
\end{equation}
where $\bar{\psi}_{KG}(\bar{X}^{\sigma})=\phi[x^{\sigma}\rightarrow x^{\sigma}_{0}+(\bar{X}^{\sigma}-1)dx^{\sigma}_{0}]$, representing the fluctuation of the $\phi$ field near $x^{\sigma}_{0}$ as driven by $\bar{X}^{\sigma}$ linear fluctuates. Thus the linear measurement conditions can be expressed as
\begin{equation}
\begin{cases}
&\frac{dL^{\alpha}}{dX^{\alpha}}=\zeta^{\alpha}(\frac{\bar{L}^{\alpha}}{L^{\alpha}}-1);\\
&\frac{d\zeta^{\alpha}}{dX^{\alpha}}=[\zeta^{\alpha}]^{2}\hat{b}_{\alpha}.
\end{cases}
\end{equation}
These conditions are the same as the ones for scaling Einstein field equations in the absence of any matter and scaling vacuum Einstein equations in the presence of a cosmological constant.

Following the same procedure, the Dirac equation \cite{bibitem18, bibitem19}
\begin{equation}
\begin{aligned}
&[i\hat{\gamma}^{\mu}\frac{\partial}{\partial x^{\mu}}-\frac{m c}{\hbar}]\psi=0\\
\end{aligned}
\end{equation}
with $\hat{\gamma}^{\mu}$ being gamma matrices, can be written in scaling form of
\begin{equation}
\begin{aligned}
&[i\hat{\gamma}^{\mu}\hat{a}_{\mu}\frac{\partial}{\partial X^{\mu}}-\frac{r_{m} r_{c}}{r_{\hbar}}\frac{[dx_{0}][m_{0}][[dx_{0}]/[dt_{0}]]}{[\hbar_{0}]}]\psi=0.
\end{aligned}
\end{equation}
After dimensional analysis, we get
\begin{equation}
\begin{aligned}
&[i\hat{\gamma}^{\mu}\hat{a}_{\mu}\frac{\partial}{\partial X^{\mu}}-\frac{r_{m} r_{c}}{r_{\hbar}}]\psi=0.
\end{aligned}
\end{equation}
Here, $\psi=\psi(x^{\sigma}\rightarrow x^{\sigma}_{0}+L^{\sigma}(X^{\sigma})dx^{\sigma}_{0})$.

With the linear mapping conditions of
\begin{equation}
\begin{aligned}
&\frac{\partial}{\partial x^{\alpha}}=\frac{1}{dx_{0}^{\alpha}}\frac{\partial}{\partial \bar{X}^{\alpha}}
\end{aligned}
\end{equation}
the scaling Dirac equation is
\begin{equation}
\begin{aligned}
&[i\hat{\gamma}^{\mu}\frac{\partial}{\partial \bar{X}^{\mu}}-\frac{r_{m} r_{c}}{r_{\hbar}}]\bar{\psi}_{Dirca}=0,
\end{aligned}
\end{equation}
where $\bar{\psi}_{Dirca}(\bar{X}^{\sigma})=\psi[x^{\sigma}\rightarrow x^{\sigma}_{0}+(\bar{X}^{\sigma}-1)dx^{\sigma}_{0}]$. Thus, we have
\begin{equation}
\begin{aligned}
&\frac{dL^{\alpha}}{dX^{\alpha}}=\zeta^{\alpha}(\frac{\bar{L}^{\alpha}}{L^{\alpha}}-1).
\end{aligned}
\end{equation}

The UV problem is also avoidable for the evolution of the equations (Eq. 74, 80) in spacetime described by \( \bar{X}^{\alpha} \), in the case of \( \bar{X}^{\alpha} > 1 \) (external measurements). The discussions for spacetime structures should be consistent with the previous section.

The scaling transformations of the Klein-Gordon and Dirac equations illustrate how changes in measurement scales can influence their forms. By incorporating scaling factors, this approach offers a potential step toward integrating quantum mechanics with the fluctuating geometry of spacetime. The shift from nonlinear to linear measurements introduces scaling factors, which impact the field equations. These transformations suggest that spacetime may possess a scaling nature, which could affect particle dynamics, thereby providing a new perspective on the relationship between quantum fluctuations and spacetime geometry. This approach may offer fresh insights into fundamental constants and quantum field theory, suggesting a possible connection between measurement scales and fundamental physics. However, further exploration and empirical validation are required to fully understand these implications.

\section{Summary}
The Planck length represents the smallest resolution in physics, where the quantum fluctuations of spacetime become evident, as inferred by the uncertainty principle. This article introduces a novel concept of spacetime microelement measurements with scaling factors, establishing a framework to explore the scaling behavior of spacetime at the Planck scale. In this framework, the microstructure of spacetime fluctuations can be characterized by three types which are stable, linear fluctuation, and non-linear fluctuations incorporated with a scaling function.

A scaling-characterized metric tensor is derived from the Lorentz scalar line element, which is expressed through micro measurements in the form of scaling functions. By applying local Lorentz transformations to these micro measurements, the spacetime structure coefficients can be defined, constraining the dimensions and components of spacetime in the rest frame. Besides, the scaled Lorentz factor at the micro-scale can be obtained from the constraint of them.

Key physical differential equations such as geodesic, Einstein field, Klein-Gordon, and Dirac can be transformed into scaling form, reflecting the localized spacetime dynamics properties at the micro-scale. Under stable conditions, spacetime can revert to a classical non-fluctuating state, with a symmetry of scale invariance by analyzing the renormalization group equations for micro measurements, corresponding to the fields and functions defined in the classical spacetime being stable. These equations acquire new forms characterized by the linear scale measurements $\bar{X}^{\alpha}$, within the mapping conditions on the scale transformation factors $\zeta^{\alpha}$. In classical spacetime, differential equations capture the local properties of fields or spacetime functions, providing well-defined values at every point without fluctuations. In classical field theory, fields have clear, fixed values throughout spacetime. However, when spacetime experiences fluctuations, these values can become variable, driven by the spacetime fluctuations themselves. Near a given point $x^{\alpha}_{0}$ in spacetime, these functions depend on the spacetime fluctuations of $X^{\alpha}$. Once the spacetime fluctuations are linearly mapped and with given forms of measurements, these fields or functions can be defined at discrete points in spacetime, resulting in well-defined values. This indicates that the linear mapping constraints with given forms of measurements can help stabilize the fluctuations while leading to discretizing spacetime. Physics might be defined by fluctuating spacetime for the forms solved by the linear mapping constraints without preset forms.

It can be summarized that the mapping conditions yield three distinct types of constraints: (1) zeroth-order constraints, as given by Equation (68); (2) first-order constraints, as delineated by Equation (53); and (3) second-order constraints, which are represented by the second set of equations in Equation (65). Here, the term 'order' refers to the order of differentiation of scaling functions $L^{\alpha}(X^{\alpha})$ with respect to scaling factors $X^{\alpha}$ and also corresponds to the mapping orders between $x^{\alpha}$ and $\bar{X}^{\alpha}$. Equations (59) and (60) represent the mapping constraints for proper measurements, connecting the coordinate-based measurements. 

It is interesting to note that the golden ratio naturally emerged in the microscopic scale of linear measurements ($L^{\alpha}=X^{\alpha}$) with the mapping constraints, suggesting a fundamental reference length that restricts micro-length measurements to specific ranges. This insight provides a potential explanation for the existence of the Planck length in addressing the problem of UV divergence. In general, for arbitrary \( L^{\alpha} = L^{\alpha}(X^{\alpha})>0 \), the UV problem is avoidable for all the equations with the linear mapping conditions in the measurements of direction towards the Plank scale. 

This work presents a coherent framework for understanding spacetime scaling at the Planck scale. While this is just the first step of a long march and these ideas are theoretically intriguing, they need further exploration and validation to assess the physical accuracy and relevance to spacetime and quantum gravity. The avoidable UV problem is at least a positive starting point. Future work will focus on identifying potential approaches to quantizing the scaling gravitational field equations in fluctuating spacetime at the Planck scale. While theoretical advancements have been made, direct experimental validation remains a challenge. Further research will explore the connections between this framework and established theories like string theory and loop quantum gravity. Indirect evidence for scaling structures might be pursued through high-energy physics experiments or cosmological observations, offering promising pathways to test the theoretical predictions. 

\section*{Acknowledgements}
We thank Dr. Baiyang Zhang and Dr. Yang Zhou for the interesting discussions and the suggestions for text editing. This work is supported by the National Natural Science Foundation of China with Grants No. 12475138 and 12147101, the Strategic Priority Research Program of Chinese Academy of Sciences (No. XDB34000000) and the Science and Technology Commission of Shanghai Municipality (23590780100).

\end{CJK*}

\begin{thebibliography}{20}
\bibitem{bibitem1} K. Becker, M. Becker, and J. H. Schwarz,
String theory and M-theory: A Modern Introduction,
\href{https://api.semanticscholar.org/CorpusID:118463239}{Cambridge U. Press, 12, 2006.}

\bibitem{bibitem2} C. Rovelli and F. Vidotto,
Covariant Loop Quantum Gravity: An Elementary Introduction to Quantum Gravity and Spinfoam Theory,
\href{https://api.semanticscholar.org/CorpusID:118083719}{Cambridge U. Press, 11, 2014}.

\bibitem{Gambini} R. Gambini and J. Pullin, A First Course in Loop Quantum Gravity. Oxford University Press (2011).

\bibitem{Ashtekar} A. Ashtekar and J. Lewandowski, Background independent quantum gravity: A status report. Classical and Quantum Gravity, 21(15), R53 (2004).

\bibitem{Thiemann} T. Thiemann,
LOOP QUANTUM GRAVITY,
\href{https://doi.org/10.1142/S0217751X08039980}{International Journal of Modern Physics A {\bf 23}, 1113-1129 (2008)}

\bibitem{Pullin} J. Pullin,
Recent developments in canonical quantum gravity,
\href{https://doi.org/10.1063/1.48785}{AIP Conf. Proc. {\bf 342}, 459–470 (1995)}

\bibitem{Gambini2} R. Gambini and J. Pullin,
Emergence of stringlike physics from Lorentz invariance in loop quantum gravity,
\href{https://doi.org/10.1063/1.48785}{International Journal of Modern Physics D {\bf 23}, No. 12, 1442023 (2014)}

\bibitem{Zwiebach} B. Zwiebach, A First Course in String Theory. Cambridge University Press (2009).

\bibitem{LeBihan} B. Le Bihan,
String theory, loop quantum gravity and eternalism,
\href{https://doi.org/10.1007/s13194-020-0275-3}{Euro Jnl Phil Sci {\bf 10}, 17 (2020).}

\bibitem{Vaid} D. Vaid,
Connecting Loop Quantum Gravity and String Theory via Quantum Geometry,
\href{ https://doi.org/10.1007/978-981-33-4408-2_55}{In: Behera, P.K., Bhatnagar, V., Shukla, P., Sinha, R. (eds) XXIII DAE High Energy Physics Symposium. DAEBRNS HEPS 2018 2018. Springer Proceedings in Physics, vol 261. Springer, Singapore.}

\bibitem{Klauder} J. R. Klauder,
Affine Quantum Gravity,
\href{https://doi.org/10.1142/S0218271803003967}{International Journal of Modern Physics D {\bf 12} 1769-1773 (2003)}

\bibitem{Oriti} D. Oriti,
The Bronstein Hypercube of Quantum Gravity,
\href{https://doi.org/10.1142/S0218271803003967}{In: Huggett N, Matsubara K, Wüthrich C, eds. Beyond Spacetime: The Foundations of Quantum Gravity. Cambridge University Press; 2020:25-52}

\bibitem{Berenstein} David Berenstein,
 Lessons in quantum gravity from quantum field theory,
\href{https://doi.org/10.1063/1.3531637}{AIP Conf. Proc. {\bf 1318}, 26–37 (2010)}

\bibitem{Shojai} A. Shojai, F. Shojai, and M. Golshani, 
 NONLOCAL EFFECTS IN QUANTUM GRAVITY,
\href{https://doi.org/10.1142/S0217732398003144}{Modern Physics Letters A {\bf 13} 2965-2969 (1998)}

\bibitem{Carlip} S. Carlip, Quantum Gravity in 2+1 Dimensions,
\href{https://doi.org/10.48550/arXiv.2312.12596}{arXiv:2312.12596 [gr-qc] (2023)}

\bibitem{Harlow}  D. Harlow and T. Numasawa, Gauging spacetime inversions in quantum gravity,
\href{https://doi.org/10.48550/arXiv.2309.10785}{arXiv:2309.10785 [hep-th] (2023)}

\bibitem{Pawlowski} J. M. Pawlowski and M. Reichert. Quantum Gravity from Dynamical Metric Fluctuations,
\href{https://doi.org/10.48550/arXiv.2309.10785}{arXiv:2309.10785 [hep-th] (2023)}

\bibitem{Braunstein} S. Braunstein,  M. Faizal, L. Krauss, F. Marino, and N. Shah, Analogue Simulations of Quantum Gravity with Fluids,
\href{https://doi.org/10.1038/s42254-023-00630-y}{Nat Rev Phys 5, 612–622 (2023)}

\bibitem{Giesel} K. Giesel,  H. Liu, P. Singh, and S. Weigl, Generalized Analysis of a Dust Collapse in Effective Loop Quantum Gravity,
\href{https://doi.org/10.48550/arXiv.2308.10953}{arXiv:2308.10953 [gr-qc] (2023)}

\bibitem{King} S. King, R. Roshan, X. Wang, G. White, and M. Yamazaki. Quantum Gravity Effects on Dark Matter and Gravitational Waves,
\href{https://link.aps.org/doi/10.1103/PhysRevD.109.024057}{Phys. Rev. D 109, 024057 (2023)}

\bibitem{Bajardi} F. Bajardi, and S. Capozziello, Minisuperspace Quantum Cosmology in f(Q) Gravity,
\href{https://doi.org/10.1140/epjc/s10052-023-11703-8}{Eur. Phys. J. C 83, 531 (2023)}

\bibitem{Chakraborty} T. Chakraborty, J. Chakravarty, V. Godet, P. Paul, and S. Raju, The Hilbert Space of de Sitter Quantum Gravity,
\href{https://doi.org/10.1007/JHEP01(2024)132}{ J. High Energ. Phys. 2024, 132 (2024)}

\bibitem{Saueressig} F. Saueressig, The Functional Renormalization Group in Quantum Gravity,
\href{https://doi.org/10.48550/arXiv.2302.14152}{arXiv:2302.14152 [hep-th] (2023)}

\bibitem{Bianchi} E. Bianchi and E. R. Livine, Loop Quantum Gravity and Quantum Information,
\href{https://doi.org/10.1007/978-981-19-3079-9_108-1}{In: Bambi, C., Modesto, L., Shapiro, I. (eds) Handbook of Quantum Gravity. Springer, Singapore (2023)}

\bibitem{bibitem3} J. A. Wheeler, 
Geons,
\href{https://doi.org/10.1103/PhysRev.97.511}{Phys. Rev. {\bf 97}, 511 (1955)}.

\bibitem{bibitem4} J. A. Wheeler with K. Ford,
Geons, Black Holes and Quantum Foam: A Life in Physics,
\href{}{W. W. Norton \& Company, 2, 2000}.

\bibitem{bibitem5} E. Tryon,
Is the Universe a Vacuum Fluctuation?
\href{https://doi.org/10.1038/246396a0}{Nature {\bf 246}, 396-397 (1973)}.



\bibitem{Kothawala} D. Kothawala,
Minimal length and small scale structure of spacetime,
\href{https://doi.org/10.1103/PhysRevD.88.104029}{Phys. Rev. D {\bf 88}, 104029 (2013)}.

\bibitem{Volovik} G.E. Volovik,
Dimensionless Physics: Planck Constant as an Element of the Minkowski Metric,
\href{https://doi.org/10.1134/S0021364022603013}{Jetp Lett. {\bf 117}, 240–244 (2023)}.

\bibitem{bibitem8} A. Einstein, Die Grundlage der allgemeinen Relativit$\ddot{a}$tstheorie. Annalen der Physik, 354(7), 769-822 (1916).
\bibitem{bibitem9} D. Hilbert, Die Grundlagen der Physik. Koniglichen Gesellschaft der Wissenschaften zu Gottingen, (1915).
\bibitem{bibitem10} J. B. Hartle, Gravity: An Introduction to Einstein's General Relativity. Addison-Wesley, (2003).

\bibitem{bibitem11} H. Minkowski, "Space and Time." Annalen der Physik, vol. 4, no. 3, pp. 301-312, (1908).
\bibitem{bibitem12} A. Einstein, "Zur Elektrodynamik bewegter K$\ddot{o}$rper." Annalen der Physik, vol. 17, no. 10, pp. 891-921 (1905).
\bibitem{bibitem13} H. A. Lorentz, Proceedings of the Royal Netherlands Academy of Arts and Sciences, vol. 6, pp. 809-831 (1904).

\bibitem{Lieu1} R. Lieu,  The Effect of Planck-Scale Spacetime Fluctuations on Lorentz Invariance at Extreme Speeds,
\href{https://doi.org/10.1086/340377}{ApJ 568 L67 (2002)}

\bibitem{Jacobson} T. Jacobson, S. Liberati,  and D. Mattingly,  A strong astrophysical constraint on the violation of special relativity by quantum gravity,
\href{https://doi.org/10.1038/nature01882}{Nature 424, 1019–1021 (2003)}

\bibitem{Abdo} A. A. Abdo et al.,  A limit on the variation of the speed of light arising from quantum gravity effects,
\href{https://doi.org/10.1038/nature08574}{Nature 462, 331-334 (2009)}

\bibitem{Zimdahl} W Zimdahl, Relativistic stochastic Boltzmann equation and fluctuations in general relativity,
\href{https://doi.org/10.1088/0264-9381/6/12/016}{Class. Quantum Grav. 6 1879 (1989)}

\bibitem{Maziashvili} M. Maziashvili, Quantum fluctuations of space-time,
\href{https://doi.org/10.48550/arXiv.hep-ph/0605146}{arXiv:hep-ph/0605146 (2006)}

\bibitem{Lieu2} R. Lieu, Relativity as the quantum mechanics of space-time measurements,
\href{https://doi.org/10.48550/arXiv.physics/0005034}{arXiv:physics/0005034 [physics.gen-ph] (2000)}

\bibitem{bibitem14} K. Schwarzschild (translation and foreword by Antoci S. and Loinger A.), On the gravitational field of a mass point according to Einstein's theory.  https://doi.org/10.48550/arXiv.physics/9905030.

\bibitem{Tsamparlis}  M. Tsamparlis, A. Paliathanasis,  Lie and Noether symmetries of geodesic equations and collineations,
\href{https://doi.org/10.1007/s10714-010-1054-9}{Gen Relativ Gravit 42, 2957–2980 (2010)}

\bibitem{Hackmann} Hackmann, Eva and Hartmann, Betti and L$\ddot{a}$mmerzahl, Claus and Sirimachan, Parinya, Complete set of solutions of the geodesic equation in the space-time of a Schwarzschild black hole pierced by a cosmic string,
\href{https://doi.org/10.1103/PhysRevD.81.064016}{Phys. Rev. D 81, 064016 (2010)}

\bibitem{bibitem15} J. M. Lee, Riemannian Manifolds: An Introduction to Curvature. Springer. (1997)

\bibitem{Richard} J.-P. Richard, Time-delay systems: an overview of some recent advances and open problems,
\href{https://doi.org/10.1016/S0005-1098(03)00167-5}{Automatica 39, 1667-1694 (2003).}

\bibitem{Lathrop} D. Lathrop, Nonlinear Dynamics and Chaos: With Applications to Physics, Biology, Chemistry, and Engineering,
\href{https://doi.org/10.1063/PT.3.2751}{Physics Today 68, 54-55 (2015).}

\bibitem{Astrom}K. J. \"A str\"om and R. Murray, Feedback Systems: An Introduction for Scientists and Engineers, Second Edition,
\href{https://press.princeton.edu/books/hardcover/9780691193984/feedback-systems}{Princeton University Press (2010).}

\bibitem{Stueckelberg} E. C. G. Stueckelberg, A. Petermann
La renormalisation des constants dans la th\'eorie de quanta,
\href{ https://doi.org/10.5169/seals-112426}{Helv. Phys. Acta (in French) {\bf 26}, 499–520 (1953)}.

\bibitem{Green1987} M.B. Green, J.H. Schwarz, and E. Witten, Superstring Theory, \href{https://doi.org/10.1017/CBO9781139248563}{Cambridge University Press. (1987)}

\bibitem{Schwartz} M. D. Schwartz, Quantum Field Theory and the Standard Model,
\href{https://doi.org/10.1017/9781139540940}{Cambridge University Press (2013).}

\bibitem{Weinberg} S. Weinberg, The Quantum Theory of Fields, 
\href{https://doi.org/10.1017/CBO9781139644167}{Cambridge University Press (1995).}

\bibitem{RovelliQG} C. Rovelli, Quantum Gravity, 
\href{https://doi.org/10.1017/CBO9780511755804}{Cambridge University Press (2004).}

\bibitem{bibitem16} O. Klein, Quantentheorie und f$\ddot{u}$nfdimensionale Relativit$\ddot{a}$tstheorie. Zeitschrift f$\ddot{u}$r Physik, 37(12), 895-906 (1926).
\bibitem{bibitem17} W. Gordon, Der Comptoneffekt nach der Schr$\ddot{o}$dingerschen Theorie. Zeitschrift f$\ddot{u}$r Physik, 40(1-2), 117-133 (1928).
\bibitem{bibitem18} P. A. M. Dirac, The quantum theory of the electron. Proceedings of the Royal Society of London. Series A, 117(778), 610-624 (1928).
\bibitem{bibitem19} J. D. Bjorken, S. D. Drell, Relativistic quantum mechanics. McGraw-Hill (1964).


\end{thebibliography}
\end{document}